\documentclass[pra,twocolumn,showkeywords,longbibliography,10pt]{revtex4-1}
\usepackage{graphicx}
\usepackage{dcolumn}
\usepackage{color}
\usepackage{times}
\usepackage{bm}
\usepackage{amssymb}
\usepackage{amsmath}
\usepackage{epsfig}
\usepackage{epstopdf}
\usepackage{dsfont}
\usepackage{subfigure}
\usepackage{tikz}
\usepackage[colorlinks, citecolor=blue, linkcolor=blue,urlcolor=blue]{hyperref}
\usepackage[mathscr]{euscript}
\usepackage{orcidlink}
\usepackage{comment}
\usepackage{appendix}
\usepackage{array}

\makeatletter
\renewcommand{\maketag@@@}[1]{\hbox{\m@th\normalsize\normalfont#1}}
\makeatother

\begin{document}
	\renewcommand{\thefootnote}{\fnsymbol {footnote}}
	
	\title{{Synergistic charging of a quantum battery via Dzyaloshinskii-Moriya interaction and non-Markovian noise}}

	\author{Si-Yu Zhang~\orcidlink{0009-0000-9944-3984}}
	\affiliation{State Key Laboratory of Opto-Electronic Information Acquisition and Protection Technology, School of Physics, Anhui University, Hefei
		230601,  People's Republic of China}
	
	\author{Meng-Long Song}
	\affiliation{State Key Laboratory of Opto-Electronic Information Acquisition and Protection Technology, School of Physics, Anhui University, Hefei 230601, People's Republic of China}

    \author{Hai-Tao Dong}
	\affiliation{State Key Laboratory of Opto-Electronic Information Acquisition and Protection Technology, School of Physics, Anhui University, Hefei 230601, People's Republic of China}

	\author{Xue-Ke Song} 
	\affiliation{State Key Laboratory of Opto-Electronic Information Acquisition and Protection Technology, School of Physics, Anhui University, Hefei 230601,  People's Republic of China}

	\author{Liu Ye} 	\affiliation{State Key Laboratory of Opto-Electronic Information Acquisition and Protection Technology, School of Physics, Anhui University, Hefei
		230601,  People's Republic of China}
	
	\author{Dong Wang~\orcidlink{0000-0002-0545-6205}} \email{dwang@ahu.edu.cn}
	\affiliation{State Key Laboratory of Opto-Electronic Information Acquisition and Protection Technology, School of Physics, Anhui University, Hefei
		230601,  People's Republic of China}

	%\author{Liu Ye} %\email{yeliu@ahu.edu.cn}
	%\affiliation{School of Physics \& Optoelectronic Engineering, Anhui University, Hefei
		%230601,  People's Republic of China}

	\date{\today}
	
	\begin{abstract}

Protecting quantum batteries from environmental noise remains a challenge. Using exact Burshtein–Kubo equations, we study a two-spin quantum battery with Dzyaloshinskii–Moriya (DM) coupling under independent local longitudinal random telegraph noise (RTN). At a fixed effective noise scale, the DM-assisted slow-switching protocol yields a superadditive transient ergotropy gain and near-maximal ergotropy during the initial charging stage, while the no-DM slow-switching control reaches comparable ergotropy only in a later revival. A finite superadditive region persists at fixed RTN amplitude, with protocol-dependent location and gain. In the Ising model, the enhancement is nonmonotonic in DM strength, with intermediate coupling giving near-optimal performance. Dressed-channel analysis associates this enhancement with changes in transition frequencies and longitudinal-noise matrix elements that reduce slow-RTN spectral weight. Superadditivity also persists in the tested anisotropic XY model.

%This study investigates the role of EPR steering in characterizing the energy dynamics of quantum batteries (QBs) within wireless charging systems. Through analysis of various coupling configurations between the charging system and reservoir, we demonstrate that enhanced steering effectively tracks energy storage growth while revealing a strong correlation between charger energy and steering parameters. Notably, steering proves superior to entanglement for energy prediction purposes. Our results indicate that steering emergence not only boosts battery charging rates and storage capacity, but also serves as a reliable indicator of extractable work. The steering activation threshold emerges as a crucial operational parameter, governing both normal battery function and charging cost optimization. Remarkably, this steering-based energy indicator maintains its effectiveness across extended wireless charging configurations, regardless of initial system states. These findings establish a novel indicator for monitoring QB energy variations, paving the way for developing high-performance quantum energy storage devices.		
	\end{abstract}
	
	\maketitle

	\section{Introduction}
 {Energy storage and transfer on the quantum scale constitute a central topic in modern quantum thermodynamics. Current research focuses primarily on modulating direct coherent interactions within many-body quantum systems to optimize energy transmission \cite{applied1,applied2,applied3,applied4,applied5}. As an important physical paradigm, the quantum battery (QB) \cite{QB1,QB2,QB3,QB4,QB5,QB6}, first proposed by Alicki \textit{et al.} \cite{propose}, is expected to exceed the performance limits of classical batteries. The thermodynamic advantage of QBs stems from the exploitation of fundamental quantum resources \cite{qn1,qn2,qn3,qn4,qn5,qn6,qn7,qn8,qn9,Caravelli2021} ---such as quantum coherence and multipartite entanglement---to enable superextensive collective charging and maximize extractable work, formally quantified as the ergotropy. To realize these advantages, various physical models have been investigated, ranging from Dicke ensembles \cite{dick1,dick2,dick3,dick4,dick5,dick6,dick7} and spin-chain configurations \cite{spin1,spin2,spin3,spin4,spin5,spin6} to alternative advanced charging schemes \cite{other1,other2,other3,other4,other5,other6,other7} and, more recently, nonreciprocal and network-based architectures~\cite{LiuNonreciprocity2026,SongExtendedQB2026,LiuTopology2026,Cong2026}, with theoretical frameworks increasingly corroborated by advanced experiments \cite{plat1,plat2,plat3,plat4}. However, the performance of realistic QBs is fundamentally constrained by their unavoidable coupling to external environments. Operating inherently as open quantum systems, they inevitably suffer from non-unitary decoherence and energy dissipation \cite{disspation1,disspation2,disspation3,disspation4,disspation5,disspation6}, which severely degrade the transient charging dynamics and cause a severe loss of the stored ergotropy. Consequently, developing mechanisms to mitigate
decoherence and understanding how structured environmental memory
modifies the charging dynamics have become important for designing
robust open QBs.

To mitigate the aforementioned noise-induced degradation, modulating internal spin exchange couplings and exploiting environmental memory effects are two primary theoretical routes. On the one hand, the Dzyaloshinskii--Moriya (DM) interaction, an antisymmetric exchange coupling, has been shown to preserve quantum correlations and enhance extractable work in spin batteries \cite{Zhang2024}. On the other hand, non-Markovian dynamics can exhibit information backflow and improve charging performance in suitable open-system settings \cite{Breuer2009,Bylicka2017,Xu2026}. DM-assisted charging in spin batteries has been
studied in Ref.~\cite{Zhang2024}, whereas non-Markovian-assisted
charging and non-Markovian protection in spin systems have been
investigated in Refs.~\cite{Xu2026} and
\cite{Benabdallah2025}, respectively. {We investigate whether near-ideal transient ergotropy can be retained under localized RTN and whether the joint DM–RTN contribution is superadditive relative to matched open-system controls. Our focus is on charging under noise, rather than surpassing the ideal noiseless performance of Ref.~\cite{Zhang2024}.
 We choose RTN because it admits an exact
Burshtein-Kubo treatment \cite{Daffer2004,Kubo1963}. The dependence
of non-Markovian reduced dynamics on the microscopic bath and coupling
structure is illustrated, for example, by spin-bath models
\cite{Krovi2007}. RTN additionally provides a standard description of
localized bistable fluctuators relevant to solid-state devices
\cite{Paladino2002,Paladino2014}. This setting allows us to determine whether the two mechanisms produce a superadditive transient ergotropy gain, rather than merely the sum of their separate effects. Such a superadditive outcome cannot be assumed a priori because the DM interaction changes the dressed transition frequencies and longitudinal-noise matrix elements that govern the RTN-induced dynamics.}

In this paper, we study transient charging of a two-spin open quantum battery under independent localized longitudinal RTN. We compare four protocols: the collective fast-switching RTN baseline without DM coupling, the DM-assisted fast-switching case, the slow-switching case without DM coupling, and the DM-assisted slow-switching case. Ergotropy and average power quantify charging performance, whereas local purity, trace distance, and population redistribution serve only as dynamical diagnostics. We map the superadditive region in the \((D,\gamma_n)\) parameter space, where \(D\) is the DM-coupling strength and \(\gamma_n\) is the RTN switching rate. To assess robustness, we perform a complementary comparison in the Ising model at fixed RTN amplitude \(b\) and examine whether the enhancement persists in an anisotropic XY model characterized by the in-plane anisotropy parameter \(\gamma_{\rm XY}\).

The remainder of this paper is organized as follows. Section~\ref{two} introduces the physical model of the open quantum battery and the exact RTN dynamics, alongside the definitions of relevant parameters and metrics. Sec.~\ref{3} investigates the dynamics of the Ising and anisotropic XY models, identifies favorable regions of superadditive enhancement, and analyzes the underlying microscopic mechanism. Sec.~\ref{4} provides the summary.}

	 \section{Model}\label{two}

\begin{figure}[htbp]
    \centering
    % ========== 第一张图：模型图 ==========
    \includegraphics[width=1.05\linewidth]{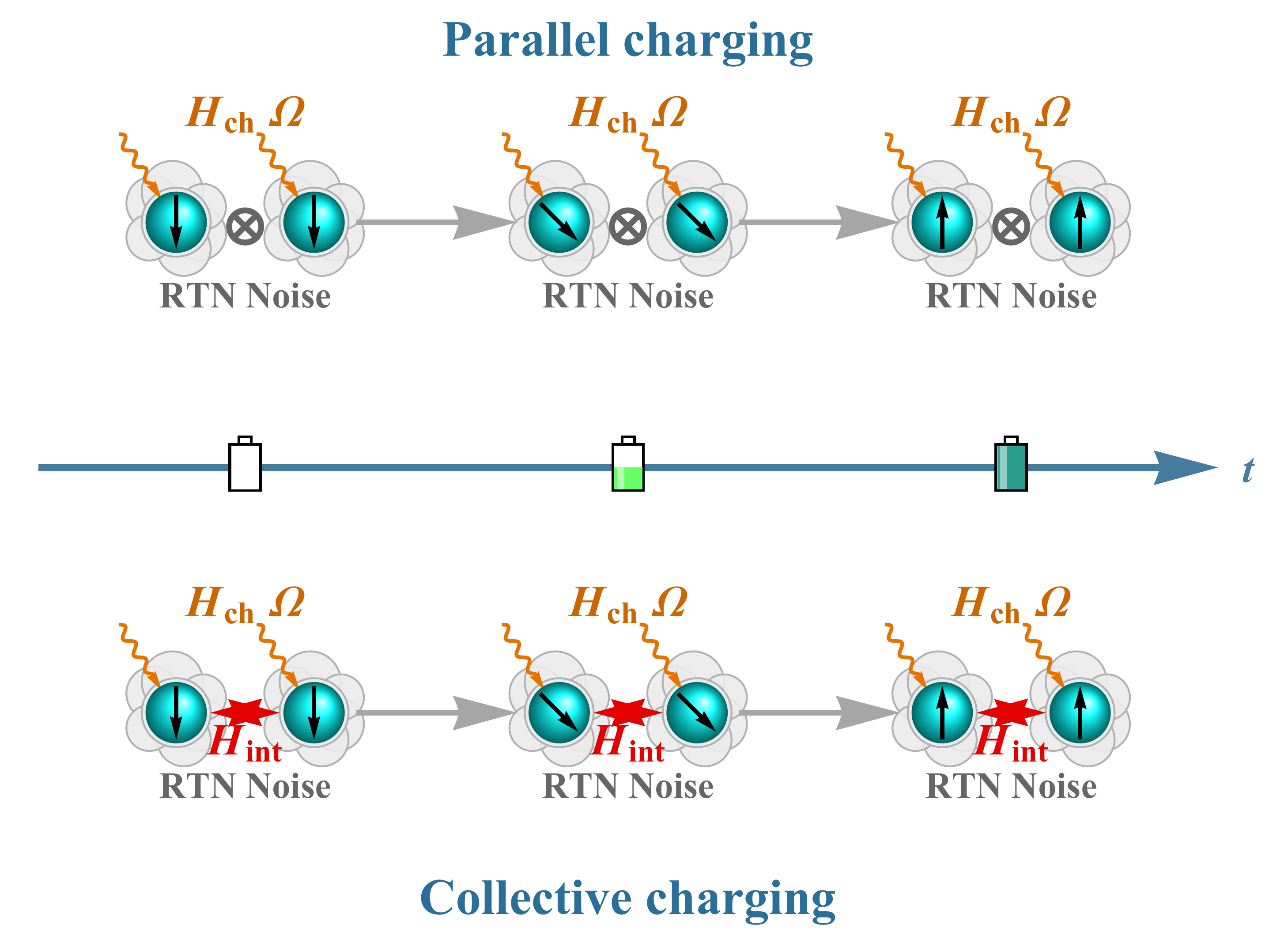}
    \caption{Schematic diagram of the two-qubit open quantum battery model under different charging schemes. The battery consists of two identical two-level systems, initialized in the fully discharged state $|\!\downarrow\downarrow\rangle$ and driven toward the fully charged state $|\!\uparrow\uparrow\rangle$ over time $t$. Both cells are independently driven by a local external charging field $H_{\rm ch}$ with strength $\Omega$, while coupled to localized random telegraph noise (RTN) environments. Parallel charging scheme: the two cells evolve independently without inter-cell interactions. Collective charging scheme: the cells interact through the internal Hamiltonian $H_{\rm int}$.}
    \label{fig1}
\end{figure}

We investigate a two-spin QB with an internal DM interaction under independent localized RTN. The battery consists of two identical two-level qubits initialized in the fully discharged state $|\!\downarrow\downarrow\rangle$. As illustrated in Fig.~\ref{fig1}, each cell is independently driven by a local coherent field and simultaneously coupled to independent localized RTN environments \cite{Paladino2002,Paladino2014}. We primarily evaluate two charging schemes: parallel charging without inter-cell interactions, and collective charging with interacting cells \cite{QB1,applied4}.

The thermodynamic energy scale of the QB is defined by its free Hamiltonian, which reads
\begin{align}
    H_0=\hbar\omega_0\sum_{n=1}^{2}\sigma_n^z .
    \label{Eq.H0}
\end{align}

Here, $\sigma_n^z$ is the spin Pauli-$z$ operator for the $n$-th spin, 
with eigenvalue equations
$\sigma_n^z|\!\downarrow\rangle=-|\!\downarrow\rangle$ and
$\sigma_n^z|\!\uparrow\rangle=|\!\uparrow\rangle$.
The parameter $\omega_0$ sets the local Zeeman energy scale, and the 
corresponding single-spin energy splitting is $2\hbar\omega_0$.
Accordingly, the ground state is $|\!\downarrow\downarrow\rangle$
with energy $-2\hbar\omega_0$, and the fully charged state is
$|\!\uparrow\uparrow\rangle$ with energy $+2\hbar\omega_0$.
Here, the local states $|\!\downarrow\rangle$ and
$|\!\uparrow\rangle$ are mapped to $|0\rangle$ and $|1\rangle$,
respectively, defining the fully charged state as $|11\rangle$.
Consequently, the adopted computational-basis convention satisfies
$\sigma_n^z|0\rangle=-|0\rangle$ and
$\sigma_n^z|1\rangle=|1\rangle$.

During charging, each cell is driven by a local coherent field. We work
directly with the following time-independent effective charging
Hamiltonian in a rotating frame:
\begin{align}
    H_{\rm ch}=\hbar\Omega\sum_{n=1}^{2}\sigma_n^x .
    \label{Eq.Hch}
\end{align}
Here, $\Omega$ is an angular frequency, and $\sigma_n^x$ is the spin
Pauli-$x$ operator for the $n$-th spin. Equations~\eqref{Eq.Hch}--\eqref{Eq.Htot}
define the effective rotating-frame model used in the calculations,
rather than the direct interaction-picture transform of a static
laboratory-frame anisotropic Hamiltonian. No additional rotating-wave
or secular approximation is imposed in the numerical propagation of
this effective model. A sufficient ideal control choice uses co-rotating
components at angular frequencies $2\omega_0$ for the single-spin
drive and $4\omega_0$ for pair excitation; its implementation is
platform dependent.

For the collective charging scheme, the internal inter-cell coupling is governed by an anisotropic Heisenberg spin-chain model incorporating the DM interaction. The interaction Hamiltonian is expressed as \cite{Zhang2024}
\begin{align}
    H_{\rm int}
    =\hbar\Big[&
    J\big((1+\gamma)\sigma_1^x\sigma_2^x
    +(1-\gamma)\sigma_1^y\sigma_2^y
    +\Delta\sigma_1^z\sigma_2^z\big) \nonumber\\
    &+D\big(\sigma_1^x\sigma_2^y
    -\sigma_1^y\sigma_2^x\big)\Big].
    \label{Eq.Hint}
\end{align}
Here, $J$ and $D$ have dimensions of angular frequency and denote, 
respectively, the nearest-neighbor exchange-coupling strength and the 
strength of the $z$-directed DM interaction. The parameters $\gamma$ 
and $\Delta$ are dimensionless and characterize the chain anisotropy. 
In the absence of the DM interaction ($D=0$), the choices 
$\gamma=1$ and $\Delta=0$ recover the conventional Ising model, whereas 
$0<\gamma<1$ and $\Delta=0$ recover the anisotropic XY model. For 
$D\neq0$, we refer to the corresponding systems as the DM-extended 
Ising and anisotropic XY models. In the parallel charging scheme, all 
internal interactions are switched off by setting $J=D=0$.

Within the effective model, $J$ and $D$ are treated as independent parameters to isolate the influence of the DM interaction. The fixed-$J$ sweeps are therefore controlled theoretical comparisons rather than a universal microscopic tuning protocol; the microscopic co-variation and accessible range of $D/J$ are platform dependent. Their implementation would require sufficiently independent control of the effective symmetric and antisymmetric exchange coefficients over the explored range.

Consequently, the total effective Hamiltonian governing the charging 
dynamics is
\begin{align}
    H_{\rm tot}=H_{\rm ch}+H_{\rm int}.
    \label{Eq.Htot}
\end{align}

To explicitly analyze the microscopic influence of the DM interaction on the battery spectrum, we express the total Hamiltonian $H_{\rm tot}$ in the ordered computational basis $\{|00\rangle,|10\rangle,|01\rangle,|11\rangle\}$. The matrix representation is given by
\begin{equation}
H_{\rm tot}
=\hbar
\begin{pmatrix}
J\Delta & \Omega & \Omega & 2J\gamma \\
\Omega & -J\Delta & 2(J+iD) & \Omega \\
\Omega & 2(J-iD) & -J\Delta & \Omega \\
2J\gamma & \Omega & \Omega & J\Delta
\end{pmatrix}.
\label{Eq.MatrixH}
\end{equation}
All entries in the matrix in Eq.~\eqref{Eq.MatrixH} therefore have 
dimensions of angular frequency. Throughout the numerical analysis, 
$\Omega$ is used as the reference frequency. Thus, the reported 
numerical values of $J$, $D$, $b$, $\gamma_n$, $\gamma_e$, and 
$\Gamma_{\rm eff}$ denote the corresponding quantities normalized by 
$\Omega$, and time is expressed through the dimensionless variable 
$\Omega t$.

Specifically, the DM interaction enters the single-excitation subspace
$\{|10\rangle,|01\rangle\}$ through the off-diagonal elements
$2(J\pm iD)$. It therefore changes both the eigenvalues and eigenvectors of the full
driven Hamiltonian. Consequently, the relevant RTN channels depend on
both the dressed transition frequencies and the corresponding matrix
elements of the longitudinal noise operators. This motivates the
spectral analysis presented below.

\subsection{RTN Environment and Exact Dynamics}

In contrast to isolated QBs governed strictly by unitary evolution, our model explicitly incorporates non-unitary dynamics induced by a stochastic environment. The RTN-driven open-system dynamics are obtained using the Burshtein--Kubo stochastic-Liouville formalism \cite{Kubo1963,Daffer2004}, which yields a closed set of equations for this classical Markov process without a weak-noise expansion.

We focus on independent local longitudinal RTN. The environmental response depends on both the noise spectrum and the system–environment coupling operator.

To characterize the intrinsic memory of the RTN source independently of the driven two-qubit Hamiltonian, we first consider exact pure dephasing of a single two-level cell. By analytically reducing the stochastic equations to the single-qubit limit (Appendix~\ref{B}), the off-diagonal element obeys $\rho_{10}(t)=\rho_{10}(0)\Lambda(t)$, where \cite{Daffer2004}
\begin{equation}
\Lambda(t)=e^{-\gamma_n t}\left[\cos(\mu t)+\frac{\gamma_n}{\mu}\sin(\mu t)\right],
\label{Eq.Decoherence}
\end{equation}
with $\mu=\sqrt{4b^2-\gamma_n^2}$. Here, $b$ is the longitudinal
system--noise coupling frequency, $\gamma_n$ is the RTN switching rate,
and the RTN correlation time is $\tau_c=(2\gamma_n)^{-1}$.
For $2b<\gamma_n$, $\mu$ is imaginary and
Eq.~\eqref{Eq.Decoherence} is understood through its equivalent
hyperbolic form.

\begin{figure}[htbp]
    \centering
    \includegraphics[width=1.0\linewidth]{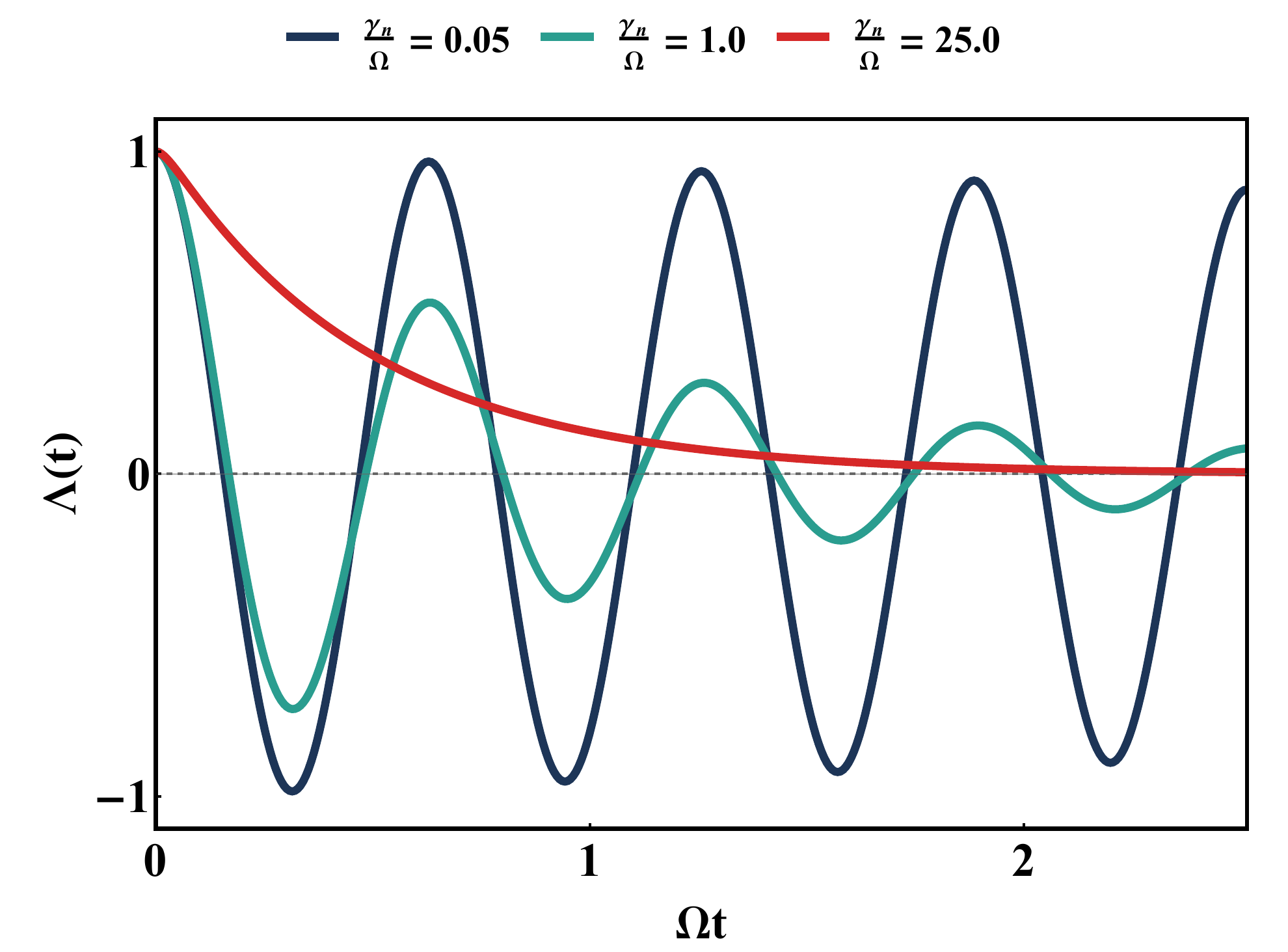}
    \caption{Exact single-qubit decoherence function $\Lambda(t)$ versus scaled time $\Omega t$ for several dimensionless RTN switching rates $\gamma_n/\Omega$, with $b/\Omega=5.0$. Small switching rates produce damped oscillations. For this pure-dephasing channel, information backflow is identified by intervals in which $|\Lambda(t)|$ increases, not by the sign of $\Lambda(t)$ itself.}
    \label{fig2}
\end{figure}

Figure~\ref{fig2} illustrates the intrinsic memory behavior of the single-qubit dephasing channel at fixed $b/\Omega$. In the slow-switching regime, $\Lambda(t)$ is oscillatory and may change sign. A negative value of $\Lambda(t)$ alone is not evidence of information backflow. For this channel, the optimal trace distance is proportional to $|\Lambda(t)|$, so BLP backflow occurs only on intervals satisfying $d|\Lambda(t)|/dt>0$. Decreasing $\gamma_n$ increases the correlation time and strengthens the nonmonotonic revivals of $|\Lambda(t)|$. In the fast-switching regime, $|\Lambda(t)|$ decays monotonically, corresponding to the Markovian pure-dephasing limit.

Figure 2 fixes \(b/\Omega=5.0\) only to illustrate single-qubit dephasing. The battery simulations instead use the comparison scale \(\Gamma_{\rm eff}/\Omega=b^2/(\gamma_n\Omega)=0.8\), giving \(b/\Omega=2.0\) and \(0.2\) for \(\gamma_n/\Omega=5.0\) and \(0.05\), respectively. The corresponding single-qubit crossover is \(\gamma_n/\Omega=3.2\), but outside the fast-switching limit \(\Gamma_{\rm eff}\) is not interpreted as a unique decay rate. We therefore call these the fast- and slow-switching RTN protocols: in the auxiliary pure-dephasing problem they produce monotonic and nonmonotonic \(|\Lambda(t)|\), respectively, whereas the BLP character of the driven two-qubit dynamics is assessed independently in Fig. 8. A complementary fixed-\(b\) comparison is presented in Fig. 9.

We model the RTN environment as independent classical stochastic fields acting locally on each spin along the $z$ axis. The two localized RTN fields are assumed to be independent and
spatially uncorrelated, modeling separate localized defect
fluctuations. During charging, the stochastic Hamiltonian is \cite{Paladino2002,Daffer2004}
\begin{align}
    H(t)=H_{\rm tot}+{\hbar}b\sum_{n=1}^{2}\nu_n(t)\sigma_n^z .
    \label{Eq.Hstochastic}
\end{align}
Here, $\nu_n(t)\in\{+1,-1\}$ are independent random telegraph variables, each switching at rate $\gamma_n$, with $\langle\nu_n(t)\rangle=0$ and $\langle\nu_n(t)\nu_m(0)\rangle=\delta_{nm}e^{-2\gamma_n t}$. The labels $+$ and $-$ below denote the instantaneous values of these classical stochastic variables; they are not quantum states of an additional environment. The free Hamiltonian $H_0$ is the thermodynamic reference in the adopted rotating frame and is not added again to the effective dynamical generator.

The exact QB dynamics are obtained from conditional density matrices $\rho_{\nu_1\nu_2}(t)$, where $\nu_1,\nu_2\in\{+,-\}$ label the four instantaneous classical noise configurations. Their temporal evolution is (see Appendix~\ref{A})
\begin{align}
    \dot{\rho}_{\nu_1\nu_2}(t)={}&-{\frac{i}{\hbar}}[H_{\nu_1\nu_2},\rho_{\nu_1\nu_2}(t)]-2\gamma_n\rho_{\nu_1\nu_2}(t)\nonumber\\
    &+\gamma_n\rho_{-\nu_1,\nu_2}(t)+\gamma_n\rho_{\nu_1,-\nu_2}(t),
    \label{Eq.BurshteinKubo}
\end{align}
with
\begin{equation*}
H_{\nu_1\nu_2}=H_{\rm ch}+H_{\rm int}+{\hbar}b(\nu_1\sigma_1^z+\nu_2\sigma_2^z).
\end{equation*}

For an initially unbiased and uncorrelated RTN distribution, the initial conditions are $\rho_{\nu_1\nu_2}(0)=\frac{1}{4}|\!\!\downarrow\downarrow\rangle\langle\downarrow\downarrow\!\!|$ \cite{Paladino2014}. The reduced state is obtained by summing over the classical configurations,
\begin{align}
    \rho(t)=\Lambda_t[\rho(0)]=\sum_{\nu_1,\nu_2}\rho_{\nu_1\nu_2}(t).
    \label{Eq.DynMap}
\end{align}
The resulting trace-preserving dynamical map includes both the coherent effective Hamiltonian and the ensemble-averaged RTN dephasing.

\subsection{Ergotropy and Charging Power}

To quantify energy-storage performance under RTN dephasing, we evaluate the ergotropy $\mathcal{E}(t)$ and the average charging power $P(t)$. Ergotropy is the maximum work extractable from $\rho(t)$ through cyclic unitary operations $U$ \cite{Kamin2020,Aziz2025}:
\begin{align}
    \mathcal{E}(t)=\operatorname{Tr}[\rho(t)H_0]-\min_U\operatorname{Tr}[U\rho(t)U^\dagger H_0].
    \label{Eq.ErgotropyDef}
\end{align}
The passive state is diagonal in the eigenbasis of $H_0$, with larger populations assigned to lower energies. Writing $\rho(t)=\sum_{i=0}^{3}r_i(t)|r_i(t)\rangle\langle r_i(t)|$ with $r_0\ge r_1\ge r_2\ge r_3$, and $H_0=\sum_{i=0}^{3}\epsilon_i|\epsilon_i\rangle\langle\epsilon_i|$ with $\epsilon_0\le\epsilon_1\le\epsilon_2\le\epsilon_3$, the passive state is $\sigma_\pi(t)=\sum_{i=0}^{3}r_i(t)|\epsilon_i\rangle\langle\epsilon_i|$. Hence,
\begin{align}
    \mathcal{E}(t)=\operatorname{Tr}[\rho(t)H_0]-\sum_{i=0}^{3}r_i(t)\epsilon_i,
    \label{Eq.ErgotropyEval}
\end{align}
and
\begin{align}
    P(t)=\mathcal{E}(t)/t.
    \label{Eq.ErgotropyEva}
\end{align}

In the numerical propagation, we set $\hbar=1$
internally and use $\Omega$ as the reference angular frequency;
$\hbar$ is restored in the displayed energy and power units.
Accordingly, $J/\Omega$, $D/\Omega$, $b/\Omega$,
$\gamma_n/\Omega$, $\gamma_e/\Omega$, and
$\Gamma_{\rm eff}/\Omega$ are dimensionless frequency ratios,
and time is reported as $\Omega t$. The thermodynamic energy
unit is $\hbar\omega_0$: $\mathcal{E}$ and $\eta_{\rm abs}$ are
reported in units of $\hbar\omega_0$, $P$ and the thermodynamic
rates in units of $\hbar\omega_0\Omega$, and the
population-transfer rate as $R/\Omega$. The relative synergy
factor $\eta_{\rm rel}$ is dimensionless.

To establish a dynamical reference, consider isolated parallel charging with $J=D=b=0$. The two independent qubits give $\mathcal{E}_{\parallel}(t)=\mathcal{E}_{\max}\sin^2(\Omega t)$, where $\mathcal{E}_{\max}=4\hbar\omega_0$, and first reach maximum ergotropy at $t_{\min}=\pi/(2\Omega)$. In the open system, the ergotropy generally develops a transient peak on this characteristic timescale. For each parameter set, we define
\[
t_{\rm opt}
=
\operatorname*{arg\,max}_{0<t\leq\pi/(2\Omega)}
\mathcal{E}_{\rm syn}(t).
\]
All four operational scenarios are evaluated synchronously at this
same $t_{\rm opt}$. Thus, the parameter-space maps characterize the
maximum transient performance attained within the prespecified
primary charging interval, rather than a later revival.

To separate the joint effect from the two individual changes, we define $\Delta\mathcal{E}_k(t)=\mathcal{E}_k(t)-\mathcal{E}_{\rm base}(t)$. Here, $\mathcal{E}_{\text{base}}$ is the ergotropy
of collective charging with $J\neq0$, $D=0$, under the
fast-switching RTN reference protocol. The synergy factors are defined in this work as
\begin{align}
    \eta_{\rm abs}(t)&=\Delta\mathcal{E}_{\rm syn}(t)-\Delta\mathcal{E}_{\rm DM}(t)-\Delta\mathcal{E}_{\rm NM}(t),
    \label{Eq.EtaAbs}\\
    \eta_{\rm rel}(t)&=\frac{\Delta\mathcal{E}_{\rm syn}(t)}{\Delta\mathcal{E}_{\rm DM}(t)+\Delta\mathcal{E}_{\rm NM}(t)}.
    \label{Eq.EtaRel}
\end{align}
Here, $\Delta\mathcal{E}_{\rm DM}$ denotes the
change produced by adding the DM interaction to the fast-switching
RTN baseline, $\Delta\mathcal{E}_{\rm NM}$ denotes the change
produced by replacing the fast-switching RTN with the
slow-switching RTN at $D=0$, and
$\Delta\mathcal{E}_{\rm syn}$ denotes the change when both
modifications are present.

Although ergotropy is nonnegative, the baseline-referenced gains
$\Delta\mathcal{E}_k$ can be negative. Hence, $\eta_{\rm rel}$ can be
negative or ill-conditioned; we interpret it as an enhancement multiplier
only when its denominator is positive and separated from zero. In that
domain, superadditivity requires both $\eta_{\rm abs}>0$ and
$\eta_{\rm rel}>1$.

\vskip 0.5cm

\section{Evolution of charging system}\label{3}

In this section, we evaluate the charging performance of the open quantum battery. We first analyze the DM-extended Ising model and then the DM-extended anisotropic XY model.

\subsection{Ising Model}

We first consider the Ising parent interaction
($\gamma=1$ and $\Delta=0$) supplemented by a variable DM coupling.
Under the fixed-$\Gamma_{\rm eff}$ protocol, we compare four
scenarios: (i) the fast-switching RTN collective baseline without
DM coupling ($D/\Omega=0$, $\gamma_n/\Omega=5.0$);
(ii) the DM-assisted fast-switching RTN protocol
($D/\Omega=4.5$, $\gamma_n/\Omega=5.0$);
(iii) the slow-switching RTN protocol without DM coupling
($D/\Omega=0$, $\gamma_n/\Omega=0.05$); and
(iv) the DM-assisted slow-switching RTN protocol
($D/\Omega=4.5$, $\gamma_n/\Omega=0.05$).
In all collective cases, $J/\Omega=1.0$.

\begin{figure}[htbp]
\centering
\includegraphics[width=\linewidth]{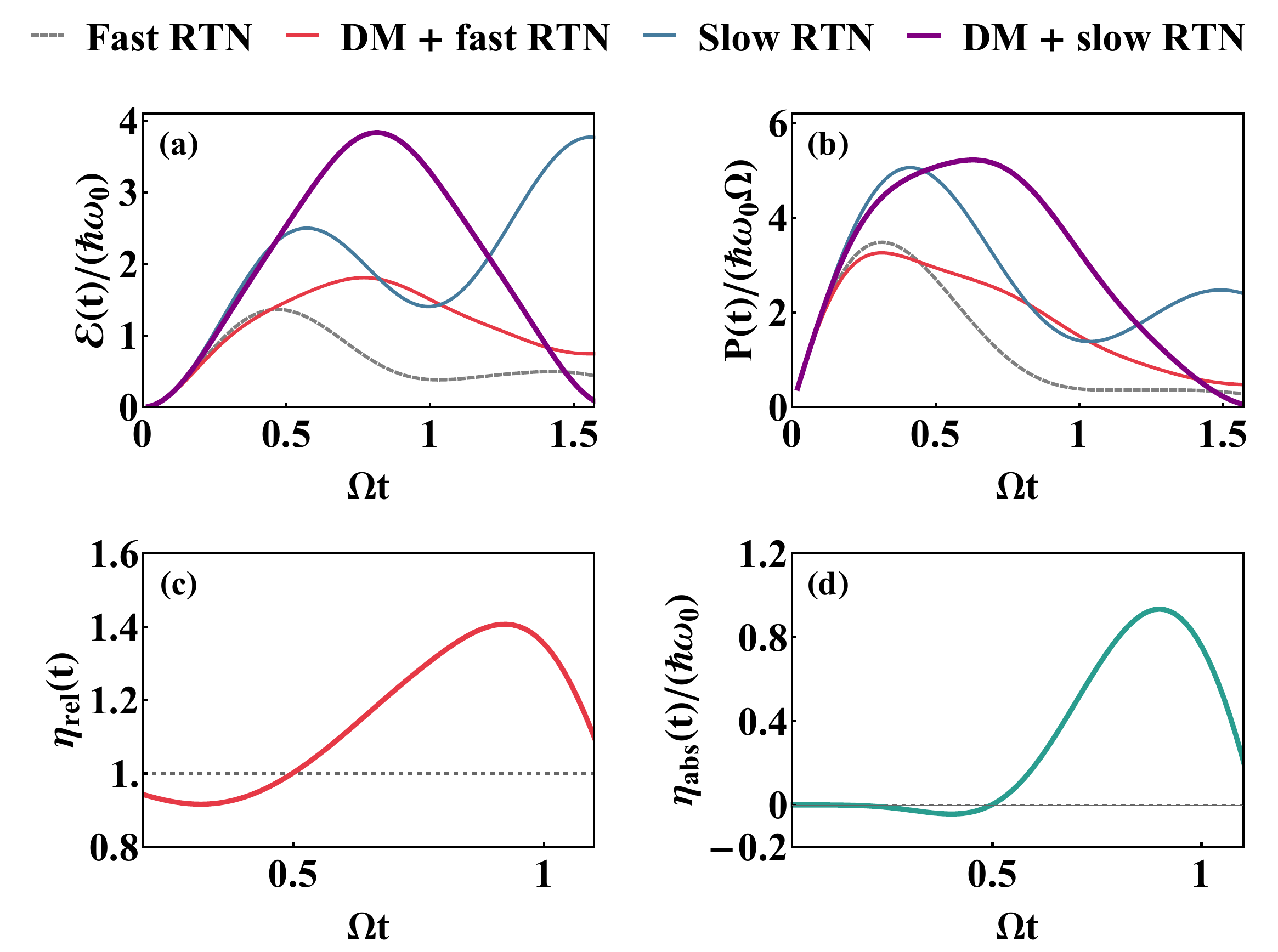}
\caption{Transient macroscopic charging dynamics and synergistic enhancement in the DM-extended Ising model ($\gamma=1$, $\Delta=0$). (a) Normalized ergotropy $\mathcal{E}(t)/(\hbar\omega_0)$ and (b) normalized average power $P(t)/(\hbar\omega_0\Omega)$ versus scaled time $\Omega t$. (c) Dimensionless relative synergy factor $\eta_{\rm rel}(t)$ and (d) absolute synergy factor $\eta_{\rm abs}(t)/(\hbar\omega_0)$. Four scenarios are compared at $J/\Omega=1.0$ and
$\Gamma_{\rm eff}/\Omega=0.8$: the fast-switching RTN collective
baseline without DM coupling ($D/\Omega=0$,
$\gamma_n/\Omega=5.0$; dashed gray); the DM-assisted
fast-switching RTN protocol ($D/\Omega=4.5$,
$\gamma_n/\Omega=5.0$; red); the slow-switching RTN protocol
without DM coupling ($D/\Omega=0$, $\gamma_n/\Omega=0.05$; blue);
and the DM-assisted slow-switching RTN protocol
($D/\Omega=4.5$, $\gamma_n/\Omega=0.05$; purple).
Values of $\eta_{\rm rel}$ for which its denominator is
nonpositive or numerically close to zero are not interpreted as
enhancement multipliers.}
\label{fig_macro}
\end{figure}

Figures~\ref{fig_macro}(a) and \ref{fig_macro}(b) show the normalized ergotropy and average power. The dashed gray fast-switching RTN collective baseline is strongly suppressed by RTN dephasing and reaches a maximum $\mathcal{E}/(\hbar\omega_0)\approx1.3$. Adding the DM interaction to this fast-switching baseline, shown by the solid red line, provides a limited ergotropy gain. The slow-switching RTN case without DM coupling, shown by the solid blue line, reaches a first peak near $\mathcal{E}/(\hbar\omega_0)\approx2.5$. The DM-assisted slow-switching protocol, shown by the solid purple line, reaches $\mathcal{E}/(\hbar\omega_0)\approx3.8$ near $\Omega t\approx0.8$ and accordingly yields a larger average power.

Although the slow-switching RTN case later develops a comparable revival near $\Omega t\approx1.6$, its first charging stage is strongly suppressed. The synergistic scheme reaches its principal ergotropy peak near $\Omega t\approx0.8$.

Figures 3(c) and 3(d) show \(\eta_{\rm rel}(t)\) and \(\eta_{\rm abs}(t)\), respectively. Before \(\Omega t\approx0.4\), \(\eta_{\rm rel}\) remains positive but below unity, while \(\eta_{\rm abs}\) is negative. When the denominator of Eq. (14) is positive and sufficiently separated from zero, these two observations indicate that the joint baseline-referenced gain is smaller than the sum of the two individual gains; they do not imply negative ergotropy. Values for which the denominator is nonpositive or numerically close to zero are not interpreted as enhancement multipliers. The nonzero commutator \([H_{\rm ch},H_{\rm int}]\neq0\) implies that the drive and the antisymmetric exchange do not generate additive dynamics and therefore permits transient cross effects. However, the commutator alone does not determine the sign of the synergy factors, which is obtained from the full stochastic evolution. In the interval \(0.6\lesssim\Omega t\lesssim1.0\), the denominator of Eq. (14) is positive, \(\eta_{\rm rel}\) rises above unity, and \(\eta_{\rm abs}\) becomes positive, indicating a superadditive transient gain.

\begin{figure}[htbp]
\centering
\includegraphics[width=1.03\linewidth]{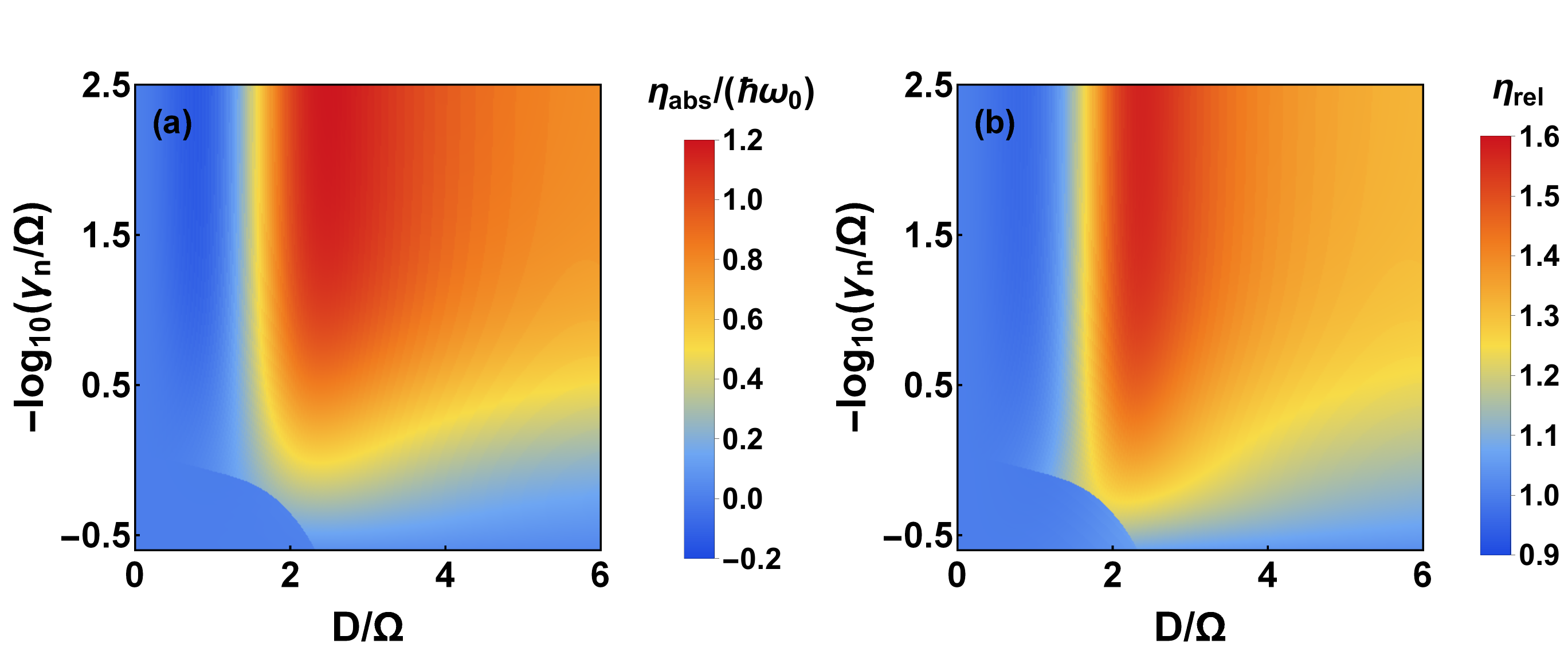}
\includegraphics[width=\linewidth]{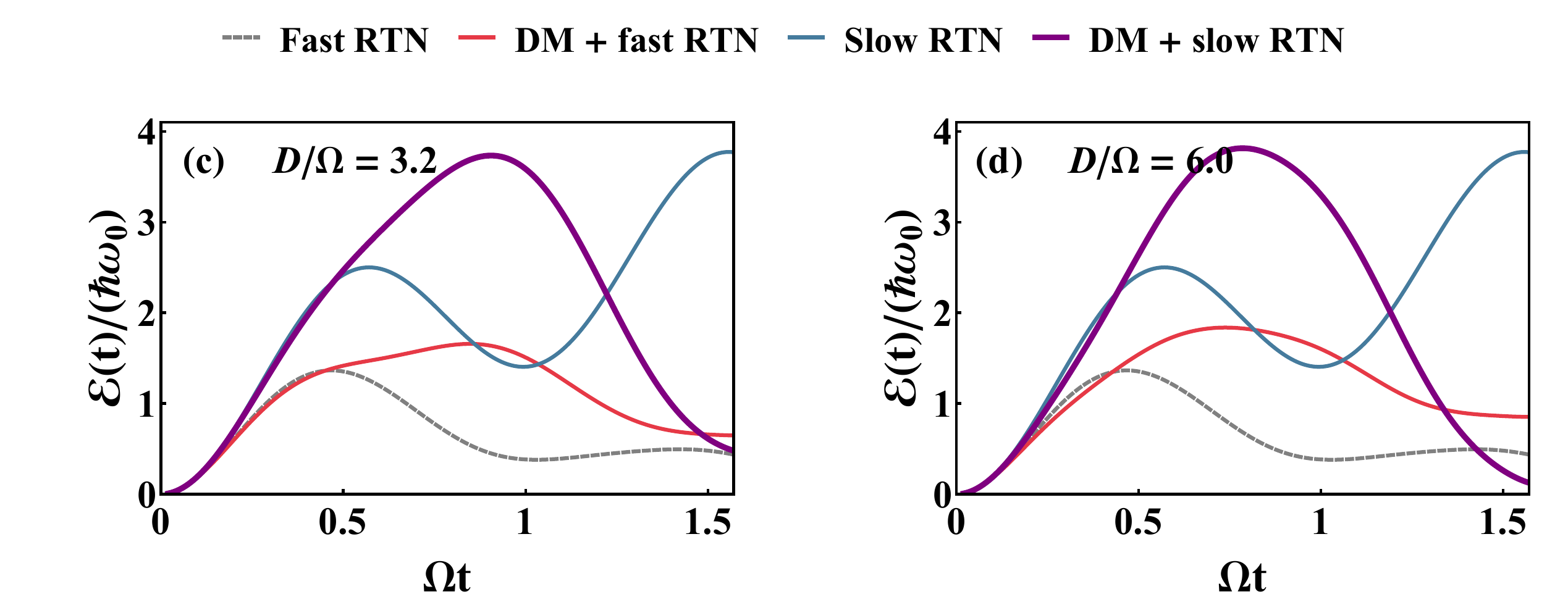}
\caption{Synergistic charging performance in the DM-extended Ising parameter space. Metrics are evaluated synchronously at
$t_{\rm opt}$, defined as the time at which the synergistic
ergotropy is maximal over the prespecified primary charging
interval $0<\Omega t\leq\pi/2$. (a) $\eta_{\rm abs}/(\hbar\omega_0)$ and (b) dimensionless $\eta_{\rm rel}$ versus $D/\Omega$ and the RTN memory (switching-rate) parameter $-\log_{10}(\gamma_n/\Omega)$, for which larger values denote slower switching and a longer RTN correlation time. The internal coupling is $J/\Omega=1.0$. (c,d) Normalized ergotropy dynamics in the slow-switching, long-memory RTN regime ($\gamma_n/\Omega=0.05$) for (c) $D/\Omega=3.2$ and (d) $D/\Omega=6.0$.}
\label{fig_heatmap_ising}
\end{figure}

To map the superadditive region, we vary $D/\Omega$ and the RTN memory parameter $-\log_{10}(\gamma_n/\Omega)$ and evaluate the metrics at $t_{\rm opt}$. This coordinate describes a property of the RTN source, not the BLP non-Markovianity of the complete reduced dynamics. Figure~\ref{fig_heatmap_ising} shows a nonmonotonic dependence on $D$.

Near $D=0$, the absolute synergy factor is negative. In this region, the small DM term perturbs the charging path but does not sufficiently alter the relevant dressed noise channels. As $D$ increases, a superadditive region emerges, and the enhancement becomes more pronounced for slower switching and longer RTN correlation time. Figures 4(c) and 4(d) show that \(D/\Omega=3.2\) already yields a peak ergotropy of approximately \(3.73\hbar\omega_0\), while increasing \(D/\Omega\) to \(6.0\) raises it only modestly to approximately \(3.81\hbar\omega_0\). Within the effective model, this identifies a moderate-ratio operating window.

At $D/\Omega=6.0$, the DM-assisted fast-switching control itself performs better, thereby increasing the individual DM contribution in the denominator and subtraction terms of Eqs.~\eqref{Eq.EtaAbs} and \eqref{Eq.EtaRel}. The resulting reduction of the synergy factors therefore does not imply a reduction of the total ergotropy. Conversely, in the fast-switching region, the contribution associated with long RTN correlation times is strongly reduced and no comparable superadditive advantage is found.

To interpret the numerical trends, consider the RTN spectral shape. From $C_\nu(\tau)=e^{-2\gamma_n|\tau|}$ and the Wiener--Khinchin relation,
\begin{equation}
S_\nu(\omega)=\frac{1}{2\pi}\int_{-\infty}^{\infty}C_\nu(\tau)e^{i\omega\tau}d\tau
=\frac{1}{\pi}\frac{2\gamma_n}{\omega^2+(2\gamma_n)^2}.
\end{equation}
Defining $\gamma_e=2\gamma_n$, this becomes
\begin{equation}
S_\nu(\omega)=\frac{1}{\pi}\frac{\gamma_e}{\omega^2+\gamma_e^2}.
\end{equation}
Because the two local RTN sources are independent, we characterize
each dressed transition using the eigenstates $|E_i(D)\rangle$ of the
full driven Hamiltonian $H_{\rm tot}$. The corresponding transition
frequency, longitudinal-noise matrix element, and spectral weight are
defined as \cite{Breuer2002}
\begin{align}
C_{ij}(D)
&=
\sum_{n=1}^{2}
\left|
\langle E_i(D)|\sigma_n^z|E_j(D)\rangle
\right|^2,
\nonumber\\
\omega_{ij}(D)
&=
\frac{E_j(D)-E_i(D)}{\hbar},
\nonumber\\
\mathcal{W}_{ij}^{(r)}(D)
&=
b_r^2
S_{\nu,r}\!\left(\omega_{ij}(D)\right)
C_{ij}(D),
\label{Eq.ChannelWeight}\\
\mathcal{W}_{\rm tot}^{(r)}(D)
&=
\sum_{i<j}
\mathcal{W}_{ij}^{(r)}(D),
\nonumber\\
\overline{\mathcal{W}}_{\rm tot}^{(r)}(D)
&=
\frac{\mathcal{W}_{\rm tot}^{(r)}(D)}
{\mathcal{W}_{\rm tot}^{(r)}(0)}.
\label{Eq.TotalChannelWeight}
\end{align}
Here, $r\in\{\mathrm{fast},\mathrm{slow}\}$ labels the two RTN
protocols, and $b_r^2=\Gamma_{\rm eff}\gamma_{n,r}$ follows the same
comparison protocol as the time-domain simulations. The quantities
$\mathcal{W}_{ij}^{(r)}$ are used as spectral-weight diagnostics
rather than exact non-Markovian decay rates.

\begin{figure}[htbp]
\centering
\includegraphics[width=\linewidth]{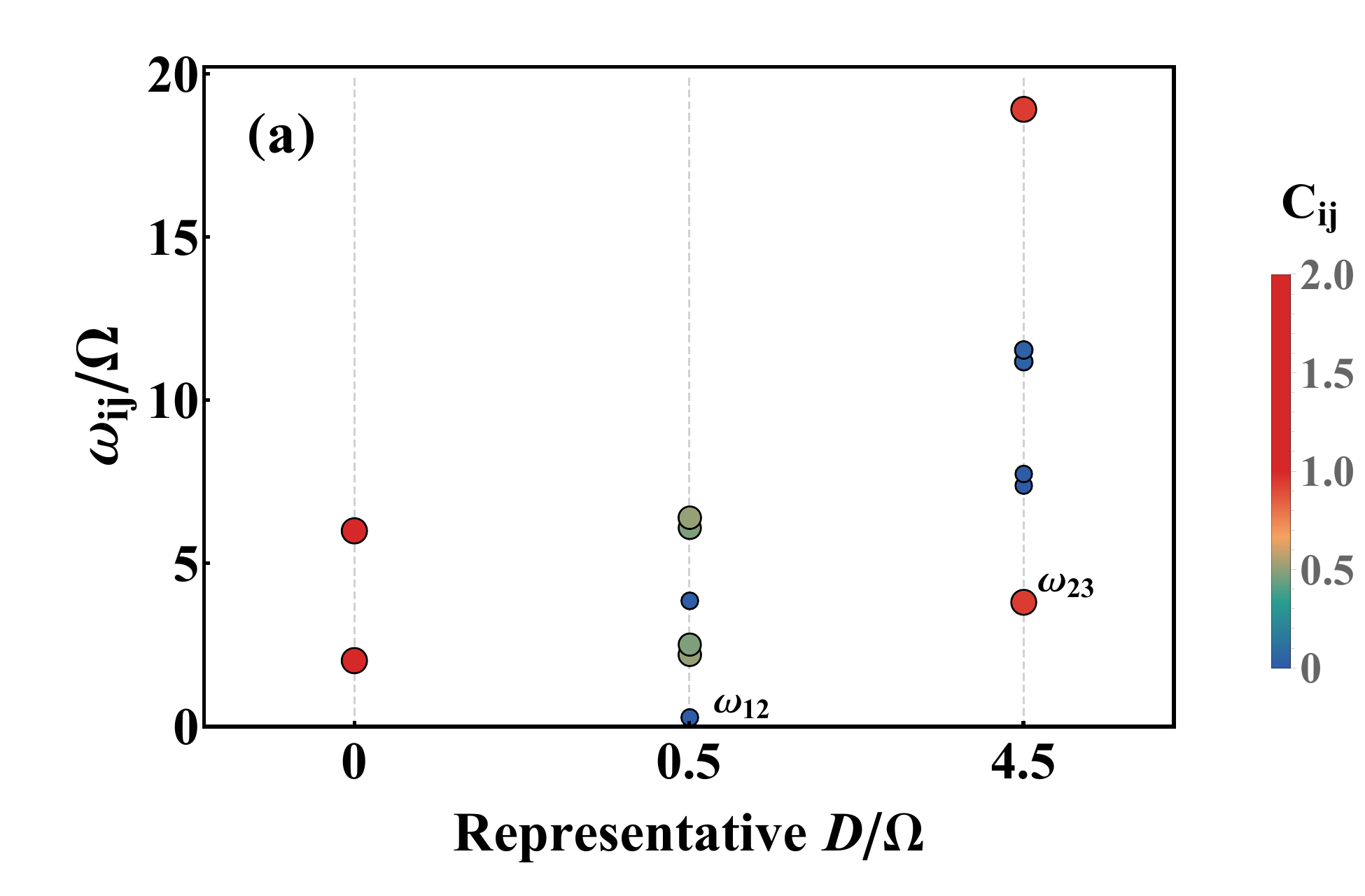}\\
\vspace{2ex}
 \includegraphics[width=1.1\linewidth]{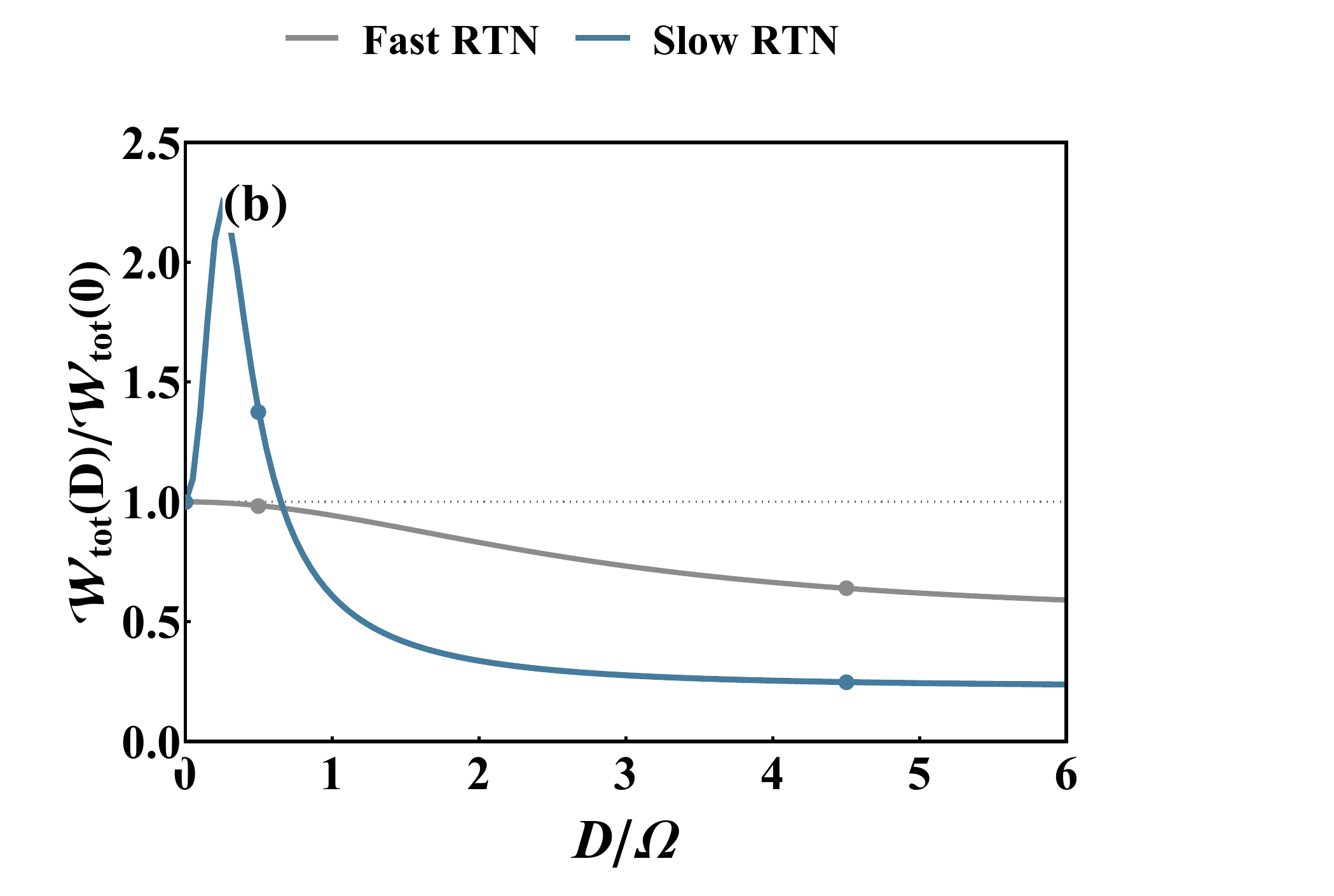}
\caption{Quantitative dressed-channel analysis for the DM-extended Ising model. (a) Positive transition frequencies $\omega_{ij}/\Omega$ of the full driven Hamiltonian at the representative values $D/\Omega=0$, $0.5$, and $4.5$. The color scale denotes $C_{ij}=\sum_{n=1}^{2}|\langle E_i|\sigma_n^z|E_j\rangle|^2$; degenerate transitions at the same frequency are grouped. (b) Normalized total dressed-channel spectral weight $\overline{\mathcal{W}}_{\rm tot}^{(r)}(D)$ for the fast- and slow-switching RTN protocols. The parameters are $J/\Omega=1$, $\Gamma_{\rm eff}/\Omega=0.8$, $\gamma_{n,\mathrm{fast}}/\Omega=5.0$, and $\gamma_{n,\mathrm{slow}}/\Omega=0.05$. The plotted weights are channel diagnostics rather than exact non-Markovian decay rates.}
\label{fig_structure}
\end{figure}

Figure~\ref{fig_structure}(a) shows the dressed transition frequencies and their longitudinal-noise matrix-element strengths. At $D/\Omega=0.5$, a low-frequency channel appears with $\omega_{12}/\Omega=0.297$ and $C_{12}=0.0345$. At \(D/\Omega=4.5\), the most strongly weighted channel is \(\omega_{23}/\Omega=3.813\), with \(C_{23}=1.892\). A higher-frequency transition has the same matrix-element factor, but the Lorentzian RTN spectrum assigns the lower-frequency \(\omega_{23}\) channel the larger spectral weight. Nevertheless, \(\omega_{23}\) still lies outside the narrow low-frequency region of the slow-RTN spectrum.

Figure~\ref{fig_structure}(b) shows the total spectral weight normalized by its value at $D=0$. At $D/\Omega=4.5$, it decreases to approximately $0.248$ for slow RTN and $0.639$ for fast RTN.

\begin{figure}[htbp]
    \centering
    \includegraphics[width=\linewidth]{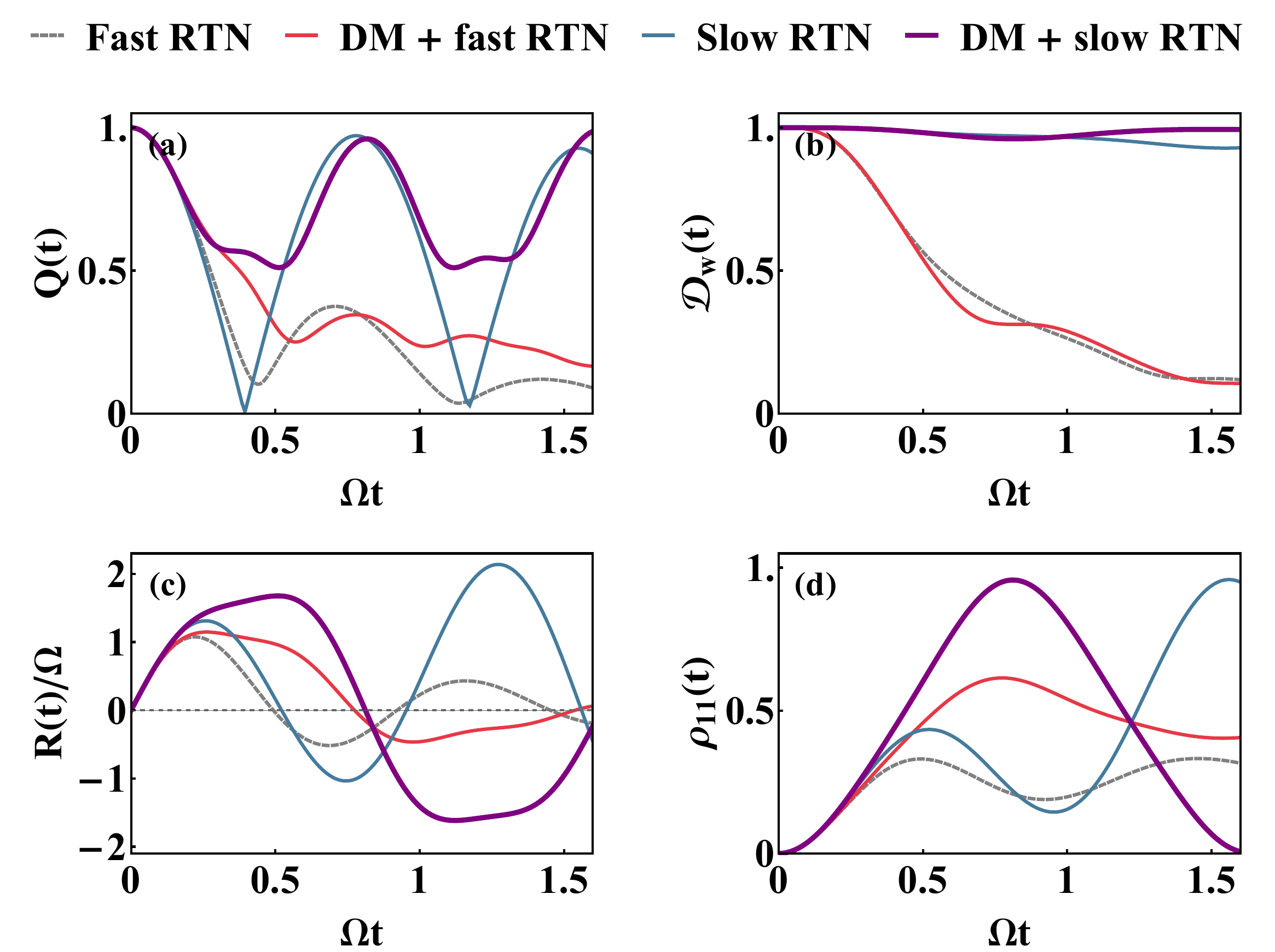}
    \caption{Transient microscopic diagnostics in the DM-extended Ising model. Temporal evolution of (a) the dimensionless two-qubit local Bloch-vector-length (local-purity) indicator $Q(t)$, (b) fixed-pair trace distance $\mathcal{D}_{\rm w}(t)$ for the orthogonal pure-state pair $|\!\downarrow\downarrow\rangle$ and $|\!\uparrow\uparrow\rangle$, (c) normalized population-transfer rate $R(t)/\Omega=\dot{\rho}_{11}(t)/\Omega$ into $|11\rangle$, and (d) fully excited-state population $\rho_{11}(t)$ for the four operational scenarios.}
    \label{fig_dynami}
\end{figure}

The boundary in Fig.~\ref{fig_heatmap_ising} displays two trends. For slow switching, \(\eta_{\rm abs}\) becomes positive only beyond a finite DM strength in the sampled parameter region.
 Toward the fast-switching region, the spectral shape broadens, and a larger $D$ is required before a comparable superadditive region appears. 
The boundary in Fig.~\ref{fig_heatmap_ising} follows the same trend as the dressed-channel weights in Fig.~\ref{fig_structure}(b): a weak DM interaction enhances a low-frequency slow-RTN channel, whereas moderate $D$ strongly reduces the total slow-RTN spectral weight.

Following the reduced-state purity construction used in
Refs.~\cite{Dong2023,Kalaga2018}, we define the following
two-qubit local-purity indicator:
\begin{equation}
Q(t)=\sqrt{\frac{Q^2(\rho_A)+Q^2(\rho_B)}{2}},
\ 
Q(\rho_k)=\sqrt{2\operatorname{Tr}(\rho_k^2)-1}.
\label{Eq.LocalPurity}
\end{equation}
For $\rho_k=(I+\mathbf{r}_k\cdot\boldsymbol{\sigma})/2$,
$Q(\rho_k)=|\mathbf{r}_k|$ measures the local Bloch-vector length.
We therefore use $Q(t)$ only as a local-purity diagnostic rather
than as a basis-dependent or resource-theoretic coherence measure.
For a globally pure two-qubit state,
$Q^2(t)=1-C^2(t)$, where $C(t)=2\sqrt{\det\rho_A(t)}$
is the concurrence. Thus, entanglement generation between the
two spins can reduce $Q(t)$ even under unitary evolution.
The generally mixed RTN-averaged state need not satisfy
this identity.

Figure~\ref{fig_dynami}(a) shows that the slow-switching RTN control without DM coupling develops a rapid reduction and later recovery of local purity. The synergistic case retains a larger local Bloch-vector length during the primary charging stage. The ergotropy, rather than $Q(t)$, remains the thermodynamic performance measure.

For two reduced evolutions from initial states $\rho_1(0)$ and $\rho_2(0)$, the trace distance is \cite{Breuer2009}
\begin{equation}
\mathcal{D}(t)=\frac{1}{2}\operatorname{Tr}\sqrt{[\rho_1(t)-\rho_2(t)]^\dagger[\rho_1(t)-\rho_2(t)]}.
\label{Eq.TraceDistance}
\end{equation}
A large, nearly constant trace distance indicates preserved
distinguishability, not information backflow. BLP backflow occurs only
when $d\mathcal{D}(t)/dt>0$.

The fully excited-state population provides a channel-resolved diagnostic. We define
\begin{align}
R(t)&\equiv\dot{\rho}_{11}(t)
=\operatorname{Tr}[|11\rangle\langle11|\dot{\rho}(t)],
\label{Eq.PopulationRate}\\
R_E(t)&\equiv\frac{d}{dt}\langle H_0\rangle
=2\hbar\omega_0[\dot{\rho}_{11}(t)-\dot{\rho}_{00}(t)],
\label{Eq.EnergyRate}\\
R_{\mathcal E}(t)&\equiv\frac{d\mathcal{E}(t)}{dt}.
\label{Eq.ErgotropyRate}
\end{align}

Here, $\rho_{aa}(t)=\langle aa|\rho(t)|aa\rangle$ ($a=0,1$).
The quantity $R(t)=\dot{\rho}_{11}(t)$ is the population-transfer
rate into $|11\rangle$, rather than the battery-energy or ergotropy
rate. In particular, $R_E(t)$ also depends on
$\dot{\rho}_{00}(t)$, while $R_{\mathcal E}(t)$ additionally contains
the time dependence of the passive-state energy.

\begin{figure}[htbp]
    \centering
    \includegraphics[width=\linewidth]{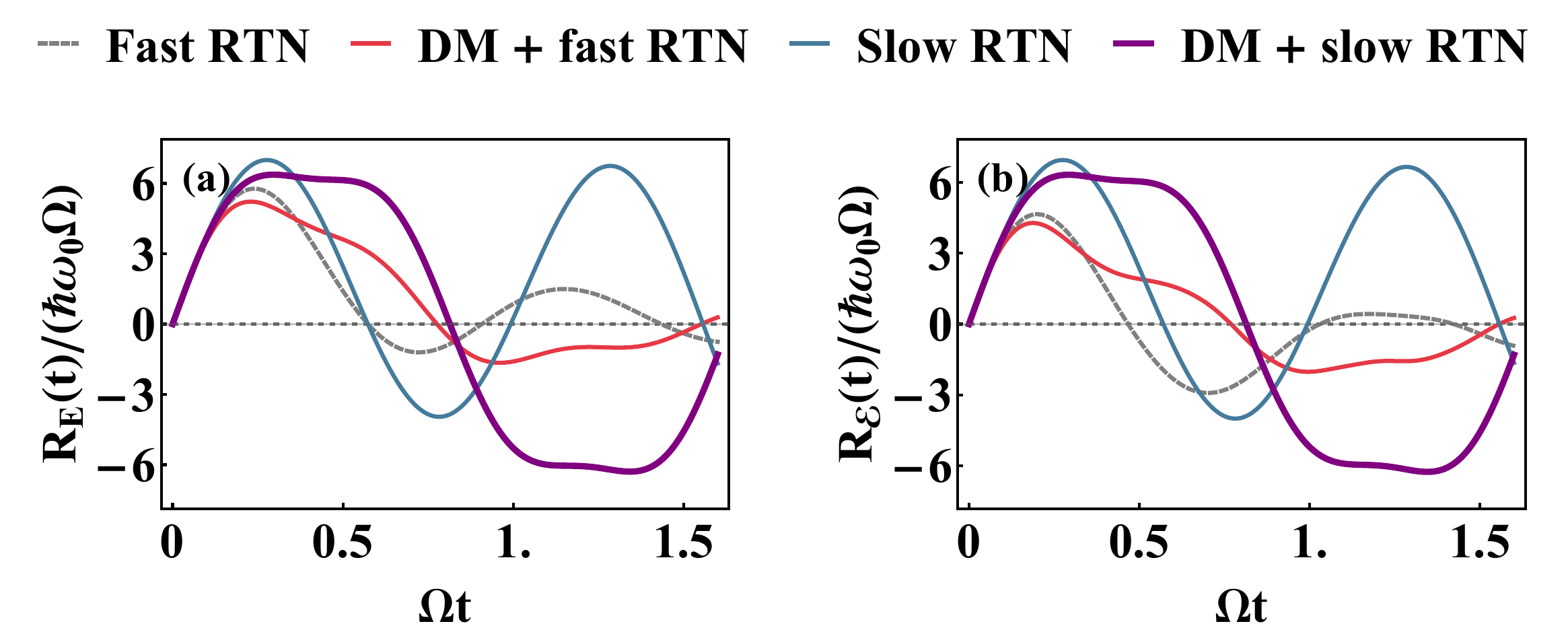}
    \caption{Normalized thermodynamic rates in the
    DM-extended Ising model. (a) Battery-energy rate
    $R_E(t)/(\hbar\omega_0\Omega)$ and (b) ergotropy rate
    $R_{\mathcal E}(t)/(\hbar\omega_0\Omega)$ versus scaled time
    $\Omega t$. The four protocols are the fast-switching RTN
    baseline, DM-assisted fast-switching RTN, slow-switching RTN
    without DM coupling, and DM-assisted slow-switching RTN.
    The parameters are $J/\Omega=1.0$,
    $\Gamma_{\rm eff}/\Omega=0.8$, and
    $D/\Omega=4.5$ for the DM-assisted protocols. Positive values
    indicate an increase in the corresponding thermodynamic
    quantity.}
    \label{fig_thermo_rates_ising}
\end{figure}

Figures~\ref{fig_dynami}(c) and \ref{fig_dynami}(d) show that the
DM-assisted slow-switching protocol sustains population transfer into
$|11\rangle$ for a longer interval and reaches a larger fully excited
population than the individual controls. To examine the corresponding
thermodynamic behavior directly,
Fig.~\ref{fig_thermo_rates_ising} shows the independently evaluated
battery-energy and ergotropy rates. Their positive initial portions
and subsequent sign changes are consistent with the buildup and local
maxima of the stored energy and ergotropy. These rates are calculated
directly and are not inferred from $\dot{\rho}_{11}(t)$.
Thermodynamic performance is therefore assessed using
$\mathcal{E}(t)$ and $P(t)$.

\begin{figure}[htbp]
    \centering
    \includegraphics[width=\linewidth]{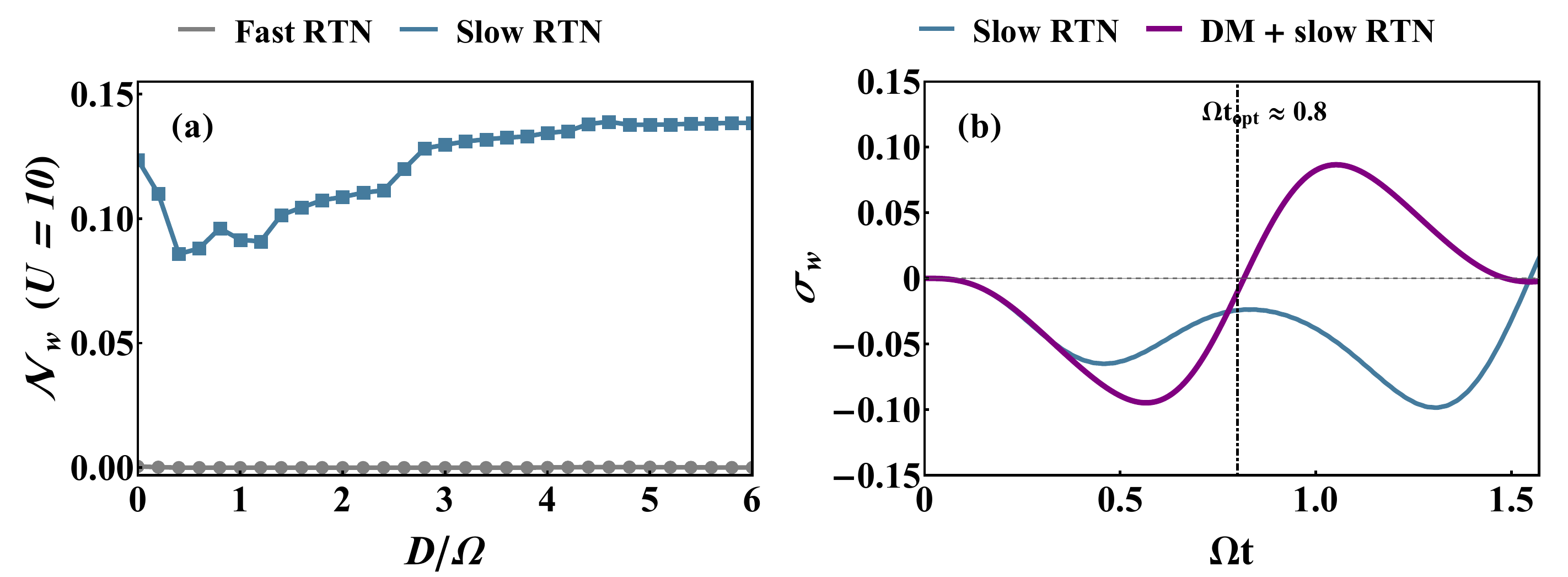}
   \caption{Finite-time trace-distance backflow
diagnostics in the DM-extended Ising model. (a) Fixed-pair backflow
witness $\mathcal{N}_{\rm w}(U=10)$ versus $D/\Omega$ for the fast-
and slow-switching RTN protocols. (b) Information-flow witness rate
$\sigma_{\rm w}(u)=d\mathcal{D}_{\rm w}(u)/du$ in the slow-switching
regime without DM coupling ($D/\Omega=0$) and with DM coupling
($D/\Omega=4.5$). Positive $\sigma_{\rm w}$ denotes trace-distance
backflow for the fixed pair, and the vertical dotted line marks the
first synergistic ergotropy maximum at
$u_{\rm opt}\simeq0.8$.}
    \label{fig_N_measure}
\end{figure}

We quantify trace-distance revivals using the BLP functional \cite{Breuer2009},
\begin{equation}
\mathcal{N}=\max_{\rho_1(0),\rho_2(0)}
\int_{\dot{\mathcal D}(t)>0}\dot{\mathcal D}(t)\,dt.
\label{Eq.BLP}
\end{equation}
In the present calculation, the maximization in
Eq.~(\ref{Eq.BLP}) is not performed. Instead, we use the same fixed
orthogonal pure-state pair
$\rho_1(0)=|\!\downarrow\downarrow\rangle
\langle\downarrow\downarrow\!|$ and
$\rho_2(0)=|\!\uparrow\uparrow\rangle
\langle\uparrow\uparrow\!|$
for every value of $D/\Omega$. The dynamics are solved independently
at each $D/\Omega$, without searching over other pure or mixed
initial-state pairs. Introducing $u=\Omega t$, we therefore define the
finite-time fixed-pair backflow witness
\begin{equation}
\mathcal{N}_{\rm w}(U)
=
\int_{0}^{U}\max\!\left\{0,\sigma_{\rm w}(u)\right\}\,du,
\
\sigma_{\rm w}(u)
=
\frac{d\mathcal{D}_{\rm w}(u)}{du},
\label{Eq.BLPWitness}
\end{equation}
where $\mathcal{D}_{\rm w}(u)$ is the trace distance for the fixed
pair. Thus, $\mathcal{N}_{\rm w}(U)$ is a finite-time lower bound on
the fully optimized BLP measure. Figure~\ref{fig_N_measure}(a) is
evaluated over the finite observation window $0\leq u\leq10$.
Full-scan numerical checks confirmed time-grid
convergence and showed that extending the finite observation window
preserves the qualitative nonmonotonic dependence on $D/\Omega$ and
the distinction between the fast- and slow-switching RTN protocols.

Figure~\ref{fig_N_measure}(a) shows a nonzero fixed-pair backflow
witness for the slow-switching RTN protocol, whereas the
fast-switching result is negligible on the scale shown.
Figure~\ref{fig_N_measure}(b) further shows that, for
$D/\Omega=4.5$, positive $\sigma_{\rm w}(u)$ occurs mainly after the
first ergotropy maximum at $u_{\rm opt}\simeq0.8$. The initial
ergotropy enhancement in Fig.~\ref{fig_macro} is therefore not
attributed to continuous BLP information backflow; trace-distance
backflow and ergotropy growth are treated as distinct quantities.

\begin{figure}[htbp]
    \centering
   \includegraphics[width=1.088\linewidth]{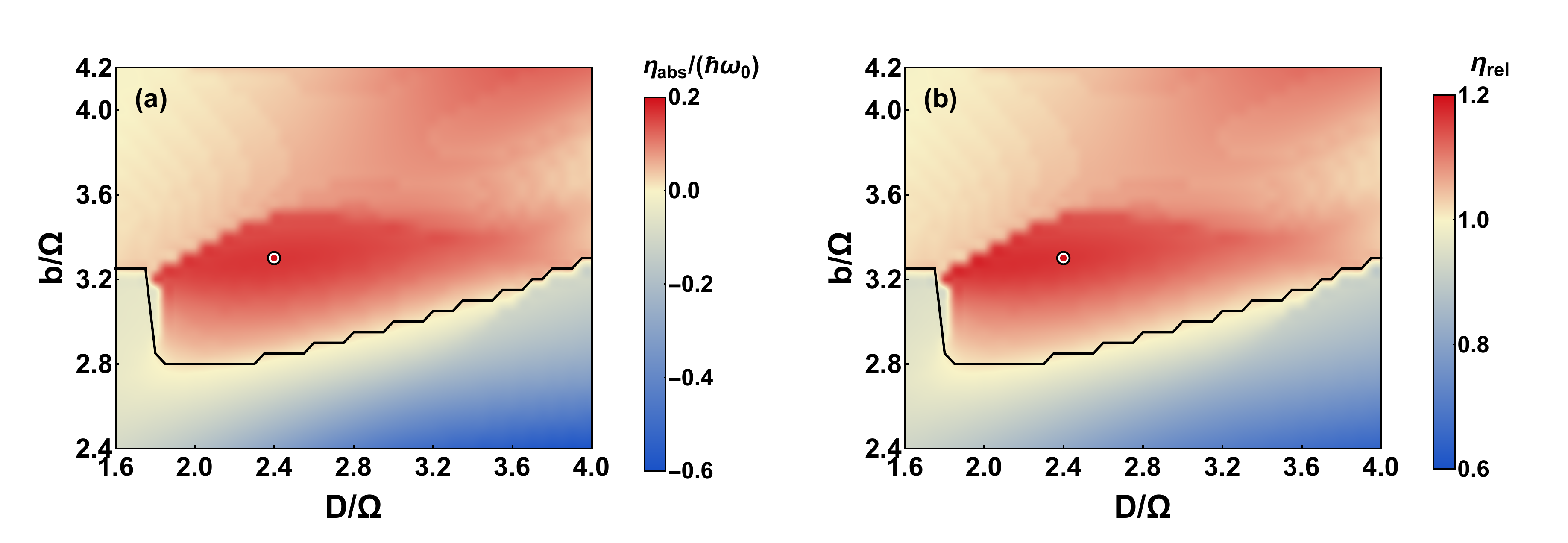}
   \includegraphics[width=\linewidth]{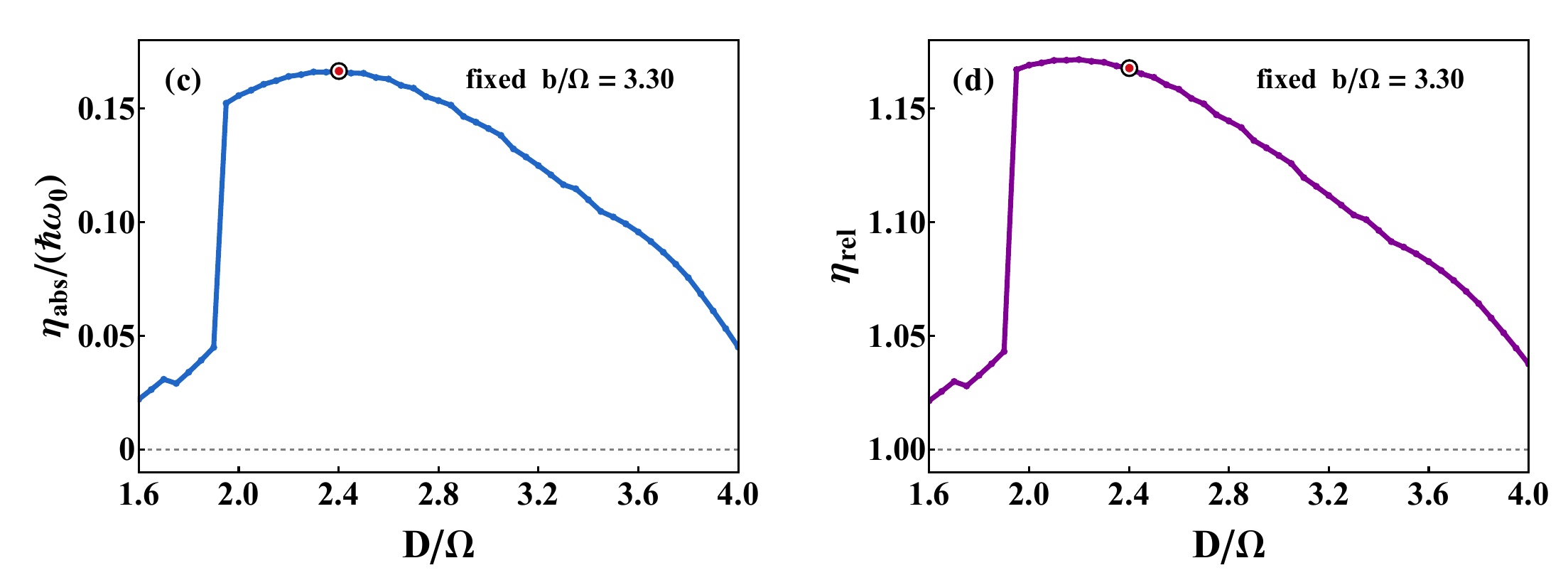}
    \caption{
    Robustness of the superadditive enhancement under the fixed-$b$
    comparison protocol in the DM-extended Ising model. The parameters
    are $J/\Omega=1.0$,
    $\gamma_{n,\mathrm{fast}}/\Omega=5.0$, and
    $\gamma_{n,\mathrm{slow}}/\Omega=0.05$, while $b/\Omega$ is held
    fixed across the four operational scenarios at each displayed
    point. The metrics are evaluated synchronously at \(t_{\rm opt}\), defined as the time at which the DM-assisted slow-switching protocol attains its maximum ergotropy over the prespecified interval \(0<\Omega t\leq\pi/2\).
    (a) $\eta_{\rm abs}/(\hbar\omega_0)$ and
    (b) $\eta_{\rm rel}$ versus $D/\Omega$ and $b/\Omega$.
    The black curve marks the lower boundary of the sampled region
    satisfying both $\eta_{\rm abs}/(\hbar\omega_0)>0$ and $\eta_{\rm rel}>1$, and the
    marked point denotes $b/\Omega=3.30$ and $D/\Omega=2.40$.
    (c) and (d) correspond to  $\eta_{\rm abs}/(\hbar\omega_0)$ and
     $\eta_{\rm rel}$ at  $b/\Omega=3.30$; the dashed
    lines mark $\eta_{\rm abs}=0$ and $\eta_{\rm rel}=1$.
    }
    \label{fig:fixed_b}
\end{figure}

Finally, we examine whether the superadditive enhancement depends on
the fixed-$\Gamma_{\rm eff}$ comparison protocol. We repeat the
Ising-model calculation using a complementary fixed-$b$ protocol. At
each point in Fig.~\ref{fig:fixed_b}, the same value of $b/\Omega$ is
used in all four operational scenarios, while
$\gamma_{n,\mathrm{fast}}/\Omega=5.0$ and
$\gamma_{n,\mathrm{slow}}/\Omega=0.05$ are retained.

A finite parameter region satisfies both $\eta_{\rm abs}>0$ and
$\eta_{\rm rel}>1$. Along the representative section
$b/\Omega=3.30$, the largest absolute gain occurs near
$D/\Omega=2.40$, where
$\eta_{\rm abs}/(\hbar\omega_0)\simeq0.166$ and
$\eta_{\rm rel}\simeq1.167$. Thus, the superadditive enhancement is
also present when the RTN amplitude is held fixed, although the location of the best-performing region within
the sampled parameter domain and the quantitative enhancement depend on
the comparison protocol.

\subsection{XY Model}

Having verified the Ising-model enhancement under both the
fixed-$\Gamma_{\rm eff}$ and fixed-$b$ comparison protocols, we next
examine its robustness against planar anisotropy. To examine whether the observed superadditive behavior persists beyond the Ising parent model, we consider an anisotropic XY parent interaction with \(\gamma_{\rm XY}=0.8\) and \(\Delta=0\), supplemented by the variable DM term.
 The fixed-$\Gamma_{\rm eff}$ environmental protocol
used in the main parameter-space comparison is kept identical to that
of the Ising case.

\begin{figure}[htbp]
\centering
\includegraphics[width=\linewidth]{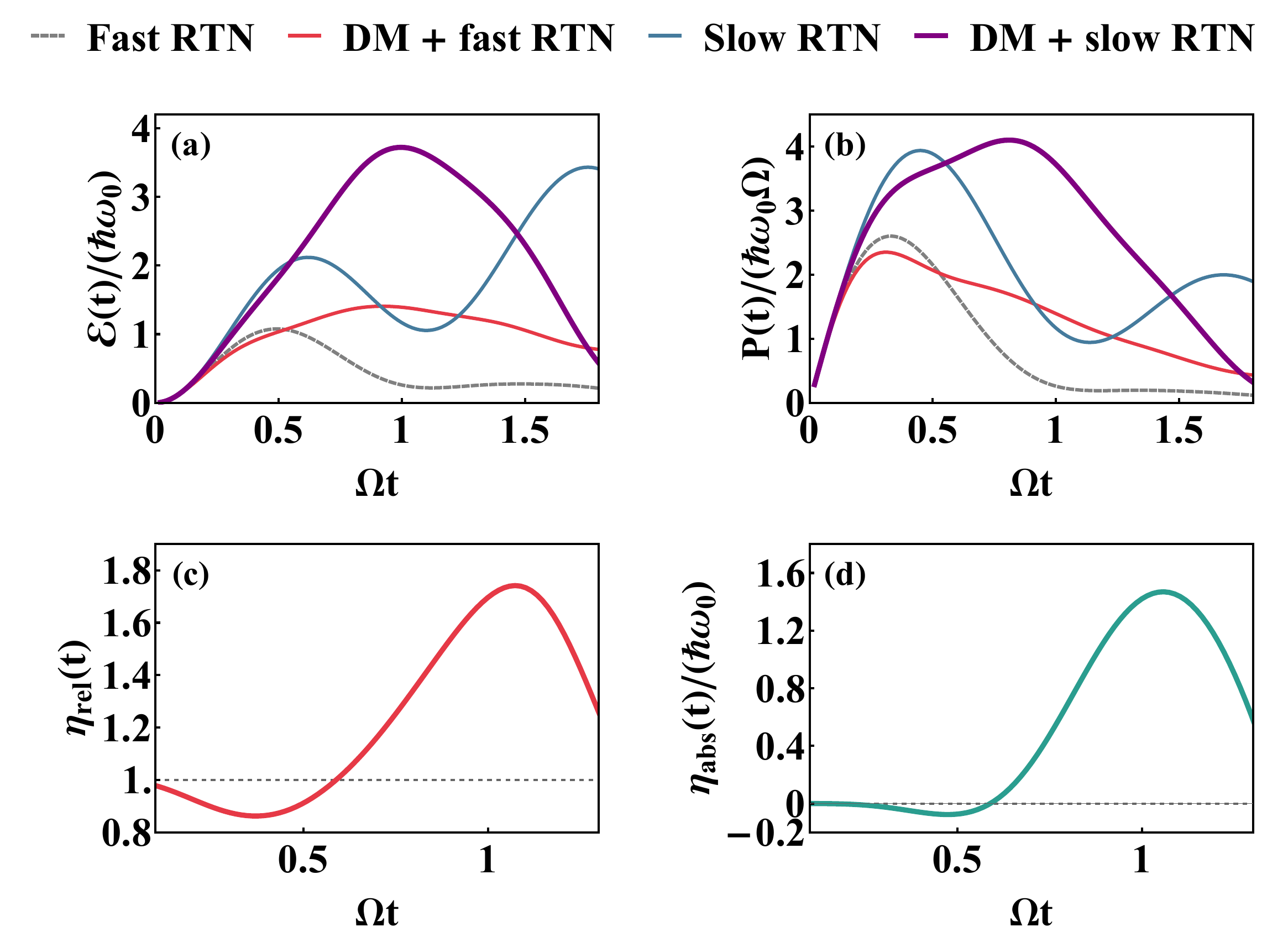}
\caption{Transient macroscopic charging dynamics and synergistic enhancement in the DM-extended anisotropic XY model with $\gamma_{XY}=0.8$ and $\Delta=0$. (a) $\mathcal{E}(t)/(\hbar\omega_0)$, (b) $P(t)/(\hbar\omega_0\Omega)$, (c) dimensionless $\eta_{\rm rel}(t)$, and (d) $\eta_{\rm abs}(t)/(\hbar\omega_0)$ versus $\Omega t$. The parameters are $J/\Omega=1.0$, $\Gamma_{\rm eff}/\Omega=0.8$, and $D/\Omega=4.0$.}
\label{fig_macro_xy}
\end{figure}

Figure~\ref{fig_macro_xy} shows the normalized charging dynamics at $D/\Omega=4.0$. The synergistic protocol retains a superadditive transient enhancement: it reaches $\mathcal{E}/(\hbar\omega_0)\approx3.7$ near $\Omega t\approx1.0$, whereas the slow-switching control without DM coupling does not attain the same high ergotropy within the
primary charging interval and exhibits its subsequent revival only at a later time. During the primary charging phase, $\eta_{\rm rel}$ exceeds unity where its denominator is positive, and $\eta_{\rm abs}$ is positive.

\begin{figure}[htbp]
\centering
\includegraphics[width=\linewidth]{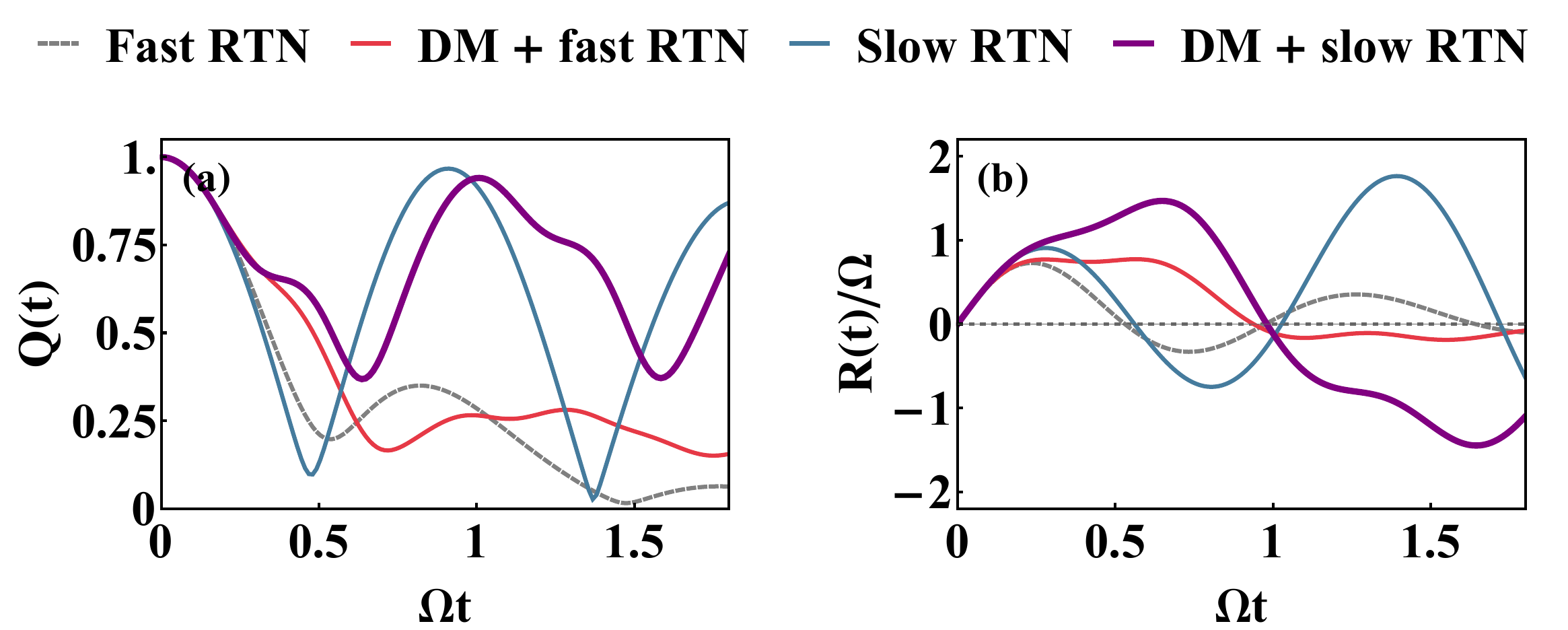}
\caption{Transient microscopic diagnostics in the DM-extended anisotropic XY model. (a) Dimensionless local Bloch-vector-length (local-purity) indicator $Q(t)$ and (b) normalized population-transfer rate $R(t)/\Omega=\dot{\rho}_{11}(t)/\Omega$ for the four operational scenarios.}
\label{fig_dynamics}
\end{figure}

\begin{figure}[htbp]
    \centering
    \includegraphics[width=\linewidth]{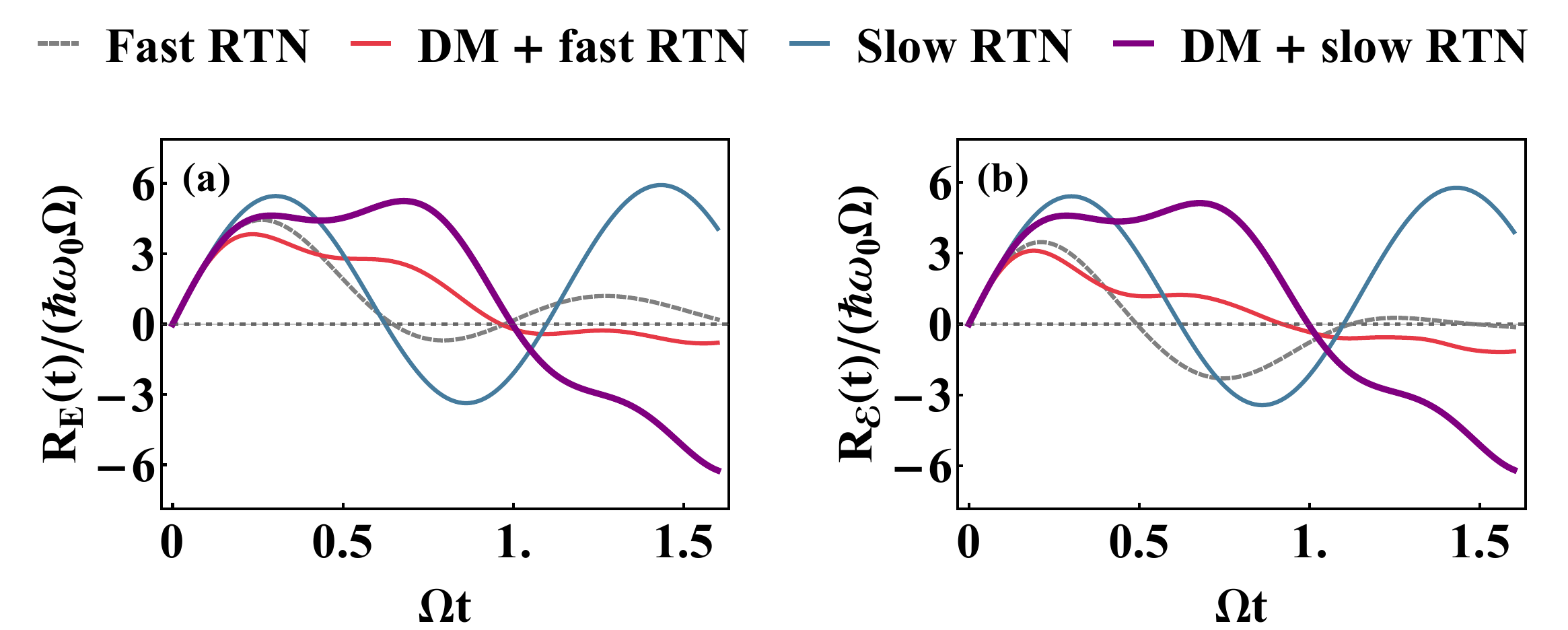}
    \caption{Normalized thermodynamic rates in the
    DM-extended anisotropic XY model with
    $\gamma_{XY}=0.8$ and $\Delta=0$. (a) Battery-energy rate
    $R_E(t)/(\hbar\omega_0\Omega)$ and (b) ergotropy rate
    $R_{\mathcal E}(t)/(\hbar\omega_0\Omega)$ versus scaled time
    $\Omega t$. The four protocols are the fast-switching RTN
    baseline, DM-assisted fast-switching RTN, slow-switching RTN
    without DM coupling, and DM-assisted slow-switching RTN.
    The parameters are $J/\Omega=1.0$,
    $\Gamma_{\rm eff}/\Omega=0.8$, and
    $D/\Omega=4.0$ for the DM-assisted protocols. Positive values
    indicate an increase in the corresponding thermodynamic
    quantity.}
    \label{fig_thermo_rates_xy}
\end{figure}

The corresponding microscopic diagnostics are shown in
Fig.~\ref{fig_dynamics}. The DM-assisted slow-switching protocol
retains a larger local Bloch-vector length and sustains population
transfer into $|11\rangle$ for a longer interval. As in the Ising
case, the population-transfer rate is not identified with a
thermodynamic charging rate. The independently evaluated energy and
ergotropy rates are shown in Fig.~\ref{fig_thermo_rates_xy}. Their initial positive intervals indicate an increase in the corresponding thermodynamic quantities, while their zero crossings identify the
local extrema of the energy and ergotropy curves. This shows that
the transient charging enhancement in the XY model remains present
when it is assessed directly through thermodynamic quantities.

\begin{figure}[htbp]
\centering
\includegraphics[width=1.03\linewidth]{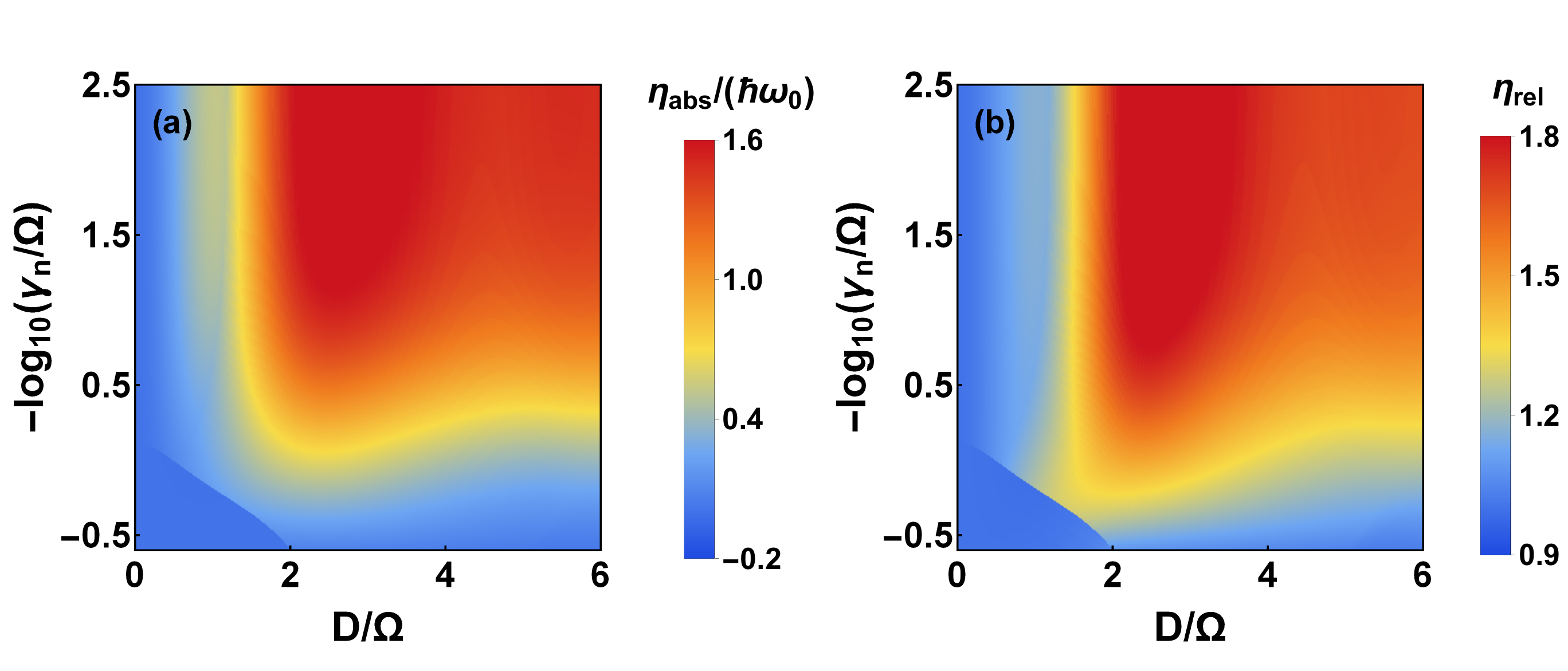}
\caption{Synergistic charging performance in the parameter space of the DM-extended anisotropic XY model ($\gamma_{XY}=0.8$, $\Delta=0$). The metrics are evaluated at $t_{\rm opt}$. (a) $\eta_{\rm abs}/(\hbar\omega_0)$ and (b) dimensionless $\eta_{\rm rel}$ versus $D/\Omega$ and the RTN memory (switching-rate) parameter $-\log_{10}(\gamma_n/\Omega)$, for which larger values indicate slower switching and a longer RTN correlation time. The internal coupling is $J/\Omega=1.0$.}
\label{fig_heatmap_xy}
\end{figure}

To map the effect under planar anisotropy, we evaluate the synergy factors as functions of $D/\Omega$ and the RTN switching parameter $-\log_{10}(\gamma_n/\Omega)$. Figure~\ref{fig_heatmap_xy} again shows a nonmonotonic dependence on $D$ and a moderate-ratio region with near-optimal transient ergotropy.

For qualitative comparison with the Ising parent model, we suppress the charging term (\(H_{\rm ch}=0\), equivalently \(\Omega=0\)) and set \(\Delta=0\). The two internal sectors then give the undriven gap combinations
\begin{equation}
\omega_{\pm}^{(0)}=
\left|2J\gamma\pm2\sqrt{J^2+D^2}\right|.
\label{Eq.UndrivenGaps}
\end{equation}

Equation~\eqref{Eq.UndrivenGaps} contains only the undriven
$H_{\rm int}$ gaps and is used for qualitative comparison. The
time-domain simulations and the channel weights are determined by the
full $H_{\rm tot}$, including $\Omega$. The shift of these internal
gaps as $\gamma$ changes from $1.0$ to $0.8$ is consistent with the
observed displacement of the numerical activation boundary, but
Eq.~\eqref{Eq.UndrivenGaps} is not used as a quantitative prediction of
that boundary.

\vskip 0.5cm

\section{Discussion and Conclusions}\label{4}

In summary, we investigated a two-qubit quantum battery under
independent localized RTN using the exact Burshtein--Kubo equations.
Under the fixed-$\Gamma_{\rm eff}$ comparison, the combined DM
interaction and slow-switching RTN yield
$\eta_{\rm abs}>0$ and $\eta_{\rm rel}>1$ over a finite parameter
region. A complementary fixed-$b$ calculation also produces a finite
superadditive region. Thus, the enhancement is not restricted to the
fixed-$\Gamma_{\rm eff}$ scaling, although the location of the
best-performing region within the sampled parameter domain and the
enhancement magnitude are protocol dependent. In the Ising model, the
combined protocol reaches a high ergotropy within the primary charging
interval and provides a larger transient average power, whereas the
slow-switching control without DM reaches a comparable ergotropy only
during a later revival.

At fixed $J/\Omega=1$, the enhancement is nonmonotonic in $D/\Omega$.
Moderate $D/\Omega$ already produces near-optimal ergotropy, while a
larger $D$ also improves the DM-assisted fast-switching control and
therefore reduces the evaluated superadditive gain. The calculated
dressed-channel weights show that the DM interaction modifies both the
transition frequencies and the longitudinal-noise matrix elements and
is accompanied, in the moderate-$D$ regime, by a stronger relative
reduction of the slow-RTN spectral weight than of the fast-RTN weight.
This provides a quantitative spectral interpretation of the observed
superadditive region.

Under the fixed-$\Gamma_{\rm eff}$ protocol, the superadditive regime
also persists in the anisotropic XY model with $\gamma_{XY}=0.8$. The present analysis concerns the charging stage and the longitudinal RTN channel; it does not establish universal decoupling from arbitrary environments. The charging field \(H_{\rm ch}\) remains explicitly active in our model, whereas coupling to a load during an extraction stage must be engineered independently.

\vskip 0.5cm

\begin{acknowledgements}
This work was supported by the National Science Foundation of China (Grant Nos.~12475009, 12075001, and 62471001), Anhui Provincial University Scientific Research Major Project (Grant No.~2024AH040008), the Anhui Provincial Natural Science Foundation (Grant No.~2508085ZD001), and Anhui Province Science and Technology Innovation Project (Grant No.~202423r06050004).
\end{acknowledgements}

\appendix
\raggedbottom

\section{Coupled Non-Markovian Master Equations}\label{A}
\label{app:master_equations}
This appendix derives the stochastic-Liouville equations used in the main text. For two independent local RTN fields along the $z$ axis,
\begin{equation}
H(t)=H_{\rm tot}+{\hbar}b\nu_1(t)\sigma_1^z+{\hbar}b\nu_2(t)\sigma_2^z.
\label{Eq.AppStochasticH}
\end{equation}
The variables $\nu_k(t)\in\{+1,-1\}$ ($k=1,2$) are independent classical stochastic processes with $\langle\nu_k(t)\rangle=0$ and $\langle\nu_k(t)\nu_{k'}(t')\rangle=\delta_{kk'}e^{-2\gamma_n|t-t'|}$.

For a fixed noise realization, the density matrix obeys
\begin{equation}
\frac{\partial}{\partial t}\rho_{\rm stoch}(t)=-{\frac{i}{\hbar}}[H(t),\rho_{\rm stoch}(t)].
\label{Eq.AppLiouville}
\end{equation}
Define the conditional marginal matrices
\begin{equation}
\rho_{\nu_1\nu_2}(t)=\left\langle\rho_{\rm stoch}(t)\,
\delta_{\nu_1(t),\nu_1}\delta_{\nu_2(t),\nu_2}\right\rangle.
\end{equation}
The four labels denote classical RTN configurations, and $\rho(t)=\sum_{\nu_1,\nu_2}\rho_{\nu_1\nu_2}(t)$. Retaining all single-flip contributions during $dt$ gives
\begin{align}
\dot{\rho}_{++}(t)
&=
-{\frac{i}{\hbar}}
[H_{++},\rho_{++}(t)]
-2\gamma_n\rho_{++}(t)
\nonumber\\
&\quad
+\gamma_n\bigl[\rho_{+-}(t)+\rho_{-+}(t)\bigr],
\label{eq:app_master_pp}
\\[0.3em]
\dot{\rho}_{+-}(t)
&=
-{\frac{i}{\hbar}}
[H_{+-},\rho_{+-}(t)]
-2\gamma_n\rho_{+-}(t)
\nonumber\\
&\quad
+\gamma_n\bigl[\rho_{++}(t)+\rho_{--}(t)\bigr],
\label{eq:app_master_pm}
\\[0.3em]
\dot{\rho}_{-+}(t)
&=
-{\frac{i}{\hbar}}
[H_{-+},\rho_{-+}(t)]
-2\gamma_n\rho_{-+}(t)
\nonumber\\
&\quad
+\gamma_n\bigl[\rho_{++}(t)+\rho_{--}(t)\bigr],
\label{eq:app_master_mp}
\\[0.3em]
\dot{\rho}_{--}(t)
&=
-{\frac{i}{\hbar}}
[H_{--},\rho_{--}(t)]
-2\gamma_n\rho_{--}(t)
\nonumber\\
&\quad
+\gamma_n\bigl[\rho_{+-}(t)+\rho_{-+}(t)\bigr].
\label{eq:app_master_mm}
\end{align}
Here,
\begin{equation}
H_{\nu_1\nu_2}
=
H_{\rm tot}
+\hbar b
\left(
\nu_1\sigma_1^z+\nu_2\sigma_2^z
\right).
\end{equation}
Summing the four equations cancels the classical transition terms and preserves $\operatorname{Tr}\rho(t)$. Because the construction is an ensemble average of unitary trajectories, the resulting map is completely positive and trace preserving.

\section{Derivation of Eq.~\eqref{Eq.Decoherence}}\label{B}
\label{app:eq6_derivation}

To obtain the single-qubit dephasing function, set the drive and all inter-cell interactions to zero. For one qubit driven only by a localized RTN source, $\rho(t)=\rho_+(t)+\rho_-(t)$ and
\begin{align}
\dot{\rho}_+(t)&=-ib[\sigma^z,\rho_+(t)]-\gamma_n\rho_+(t)+\gamma_n\rho_-(t),
\label{eq:single_plus}\\
\dot{\rho}_-(t)&=+ib[\sigma^z,\rho_-(t)]-\gamma_n\rho_-(t)+\gamma_n\rho_+(t).
\label{eq:single_minus}
\end{align}
With the convention $\sigma^z|0\rangle=-|0\rangle$ and $\sigma^z|1\rangle=|1\rangle$, one has $[\sigma^z,|1\rangle\langle0|]=2|1\rangle\langle0|$. Defining $x(t)=\rho_{10}^+(t)+\rho_{10}^-(t)=\rho_{10}(t)$ and $y(t)=\rho_{10}^+(t)-\rho_{10}^-(t)$ gives
\begin{align}
\dot{x}(t)&=-2iby(t),\label{eq:dot_x}\\
\dot{y}(t)&=-2ibx(t)-2\gamma_ny(t).\label{eq:dot_y}
\end{align}
Eliminating $y(t)$ yields
\begin{equation}
\ddot{x}(t)+2\gamma_n\dot{x}(t)+4b^2x(t)=0.
\label{eq:second_order_ode}
\end{equation}
For $2b>\gamma_n$, the roots are
\begin{equation}
r_\pm=-\gamma_n\pm i\mu,
\qquad
\mu=\sqrt{4b^2-\gamma_n^2}.
\end{equation}
For an initially unbiased RTN source, $\rho_+(0)=\rho_-(0)=\rho(0)/2$, hence $y(0)=0$ and $\dot{x}(0)=0$. The solution is
\begin{equation}
x(t)=\rho_{10}(0)e^{-\gamma_n t}
\left[\cos(\mu t)+\frac{\gamma_n}{\mu}\sin(\mu t)\right].
\end{equation}
Therefore $\rho_{10}(t)=\Lambda(t)\rho_{10}(0)$ with $\Lambda(t)$ given in Eq.~\eqref{Eq.Decoherence} \cite{Daffer2004}. In this single-qubit pure-dephasing limit, intervals of BLP information backflow correspond to increases of $|\Lambda(t)|$, not to negative values of $\Lambda(t)$.
		
	%\bibliography{v1Notes}
\end{document}